\documentclass[reprint,amsmath,amssymb,aps,pre]{revtex4-2}

\usepackage{graphicx}
\usepackage{color}
\usepackage{dcolumn}
\usepackage{floatrow}
\usepackage{bm}
\usepackage{tikz}
\usepackage{svg}
\usetikzlibrary{shapes.geometric, arrows}

\usepackage{hyperref}
\usepackage{subfiles}
\renewcommand{\eqref}[1]{Eq.~(\ref{#1})}
\newcommand{\eqsref}[1]{Eqs.~(\ref{#1})}
\newcommand{\figref}[1]{Figure~\ref{#1}}
\newcommand{\secref}[1]{Sec.~\ref{#1}}
\newcommand{\appref}[1]{Appendix~\ref{#1}}
\usepackage{placeins}

\begin{document}

\title{Multi-objective optimization for targeted self-assembly among competing polymorphs}
\author{Sambarta Chatterjee}
\affiliation{Department of Chemistry, Princeton University, Princeton, NJ 08544, USA}
\author{William M. Jacobs}
\email{wjacobs@princeton.edu}
\affiliation{Department of Chemistry, Princeton University, Princeton, NJ 08544, USA}
\date{\today}

\begin{abstract}
  Most approaches for designing self-assembled materials focus on the thermodynamic stability of a target structure or crystal polymorph.
  Yet in practice, the outcome of a self-assembly process is often controlled by kinetic pathways.
  Here we present an efficient machine learning-guided design algorithm to identify globally optimal interaction potentials that maximize both the thermodynamic yield and kinetic accessibility of a target polymorph.
  We show that optimal potentials exist along a Pareto front, indicating the possibility of a trade-off between the thermodynamic and kinetic objectives.
  Although the extent of this trade-off depends on the target polymorph and the assembly conditions, we generically find that the trade-off arises from a competition among alternative polymorphs: The most kinetically optimal potentials, which favor the target polymorph on short timescales, tend to stabilize a competing polymorph at longer times.
  Our work establishes a general-purpose approach for multi-objective self-assembly optimization, reveals fundamental trade-offs between crystallization speed and defect formation in the presence of competing polymorphs, and suggests guiding principles for materials design algorithms that optimize for kinetic accessibility.
\end{abstract}

\maketitle

\section{Introduction}

Self-assembly---the spontaneous emergence of structures with long-range order from simple building blocks---is a versatile strategy for synthesizing complex materials, but the subunit interactions must be carefully chosen to produce custom material properties~\cite{colloid_design,colloid_SA}.
\textit{Inverse design} methodologies~\cite{torquato_rev,zunger_rev} have emerged as a powerful paradigm for solving this challenging problem.
In this approach, a target microstructure with desired physical properties is proposed, and then the interactions among the building blocks are optimized to guide the self-assembly of the material.
Inverse design has been successfully applied in many settings, such as the development of DNA-programmable colloidal materials~\cite{dna_SA_rev,dna_colloid_rev} with photonic properties~\cite{pine_diamond,will_phc_dna_colloid,liu2024inverse}.
In recent years, advanced machine-learning (ML) strategies have been proposed to guide high-dimensional searches that are inherent to inverse-design problems~\cite{ml_inv_design,truskett2020,dijkstra_ml_SA}.
Techniques including maximum-likelihood learning of pair distribution functions~\cite{truskett2016}, automatic differentiation of short simulation trajectories~\cite{goodrich,jhaveri2024discovering}, gradient-free optimization algorithms for evolutionary and combinatorial searches~\cite{Sulc2020,dijkstra2022,wang_inv}, and reinforcement learning applied to simulation protocols~\cite{neuro_whitelam} have proven effective for this task, but fundamental challenges regarding the selection of optimization objectives persist.

In particular, whereas inverse design approaches have traditionally focused on minimizing the free energy of the target microstructure~\cite{torquato_rev}, kinetic pathways strongly influence the outcome of any self-assembly process~\cite{kinetic_control_SA,kinetic_control_SA_2,will_jacs,hensley2022self}.
For this reason, inverse design should aim to simultaneously ensure thermodynamic stability \textit{and} kinetic accessibility of the target microstructure.
However, these objectives are not necessarily aligned with one another, so that optimizing for thermodynamic stability may not be sufficient to ensure kinetic accessibility of a target microstructure on experimentally realistic timescales~\cite{will_jacs,hensley2022self,dna_therm_kin,therm_kin_SA}.
It is therefore essential to devise new inverse design strategies that consider both thermodynamic and kinetic objectives on equal footing.

Here we investigate the relationship between thermodynamic and kinetic objectives for a prototypical class of self-assembly problems, in which pair potentials are designed to assemble a target crystal polymorph from an initially disordered mixture.
We first introduce a new ML-guided global optimization method, \textit{active learning via Gaussian Process Regression}~\cite{an2024}, that allows us to efficiently search a high-dimensional design space of candidate pair potentials.
We then use this approach to locate the \textit{Pareto front}~\cite{pareto_def,trubiano} that characterizes the pair-potential design space with respect to a pair of thermodynamic and kinetic objectives, represented by a \textit{finite-time self-assembly yield} and a \textit{self-assembly rate}, respectively.
Pareto-optimal pair potentials, whose thermodynamic and kinetic properties lie on the Pareto front, cannot be optimized further with respect to one objective without reducing the performance of the potential with respect to the other objective.
The Pareto front is therefore of central importance for inverse design, since it reveals the extent of any trade-off between thermodynamic and kinetic optimality for the self-assembly of a target polymorph under prescribed conditions~\cite{trubiano}.
Moreover, examination of the Pareto-optimal potentials can uncover the physical mechanism underlying a trade-off between these two design objectives.

The remainder of this article is organized as follows.
In Section~II, we define the thermodynamic and kinetic objectives and show how a pair potential that was optimized to minimize the free energy of a honeycomb lattice~\cite{nmo1,nmo2} performs when self-assembly is carried out starting from various initial densities.
In Section~III, we introduce the ML-guided multi-objective optimization method to identify Pareto-optimal potentials.
To compare the performance of our method relative to free energy-based optimization, we compute Pareto-optimal potentials for different target crystal lattices, initial densities, boundary conditions, and pair-potential design spaces in Section~IV.
Then in Section~V, we examine features of the Pareto-optimal potentials themselves, showing that a trade-off between thermodynamic and kinetic objectives generically arises due to competition among competing polymorphs.
Finally, in Section~VI, we show how the shape of the Pareto front dictates the efficacy of alternative inverse-design algorithms based on optimizing ensembles of short-duration assembly trajectories.
We summarize our conclusions in Section~VII and suggest future directions for the development of improved inverse design methodologies based on our findings.

\section{Self-assembly under a thermodynamically optimized potential}
\label{sec2}

To illustrate the problem, we begin by considering the self-assembly of a two-dimensional honeycomb (HC) lattice---the two-dimensional analog of the cubic diamond lattice---using an isotropic pair potential.
The HC crystal polymorph can be stabilized by minimizing its free-energy relative to alternative crystal lattices at equilibrium, as was first demonstrated by Rechtsman \textit{et al.}\ in Ref.~\cite{nmo1}.
That work proposed a family of pair potentials comprising a (12/10) Lennard-Jones potential, a soft repulsive exponential tail, and an attractive Gaussian well,
\begin{equation}
    U_{\text{HC}}(r; \vec a) = \frac{5}{r^{12}} - \frac{a_0}{r^{10}} + a_1 e^{-a_2 r}
    - 0.4 e^{-40 \left( r-a_3 \right) ^2}.
    \label{eq:pot}
\end{equation}
$U_{\text{HC}}(r; \vec a)$ is a function of the distance between a pair of particles, $r$, which is measured in dimensionless units relative to the colloidal particle diameter, $\sigma$.
A pair potential from this family is parameterized by the vector $\vec{a} = \{ a_0,a_1,a_2,a_3 \}$.
The \textit{design space} for this family of pair potentials is thus a 4-dimensional vector space.

\begin{figure}[t!]
    \centering
    \includegraphics[width=\columnwidth]{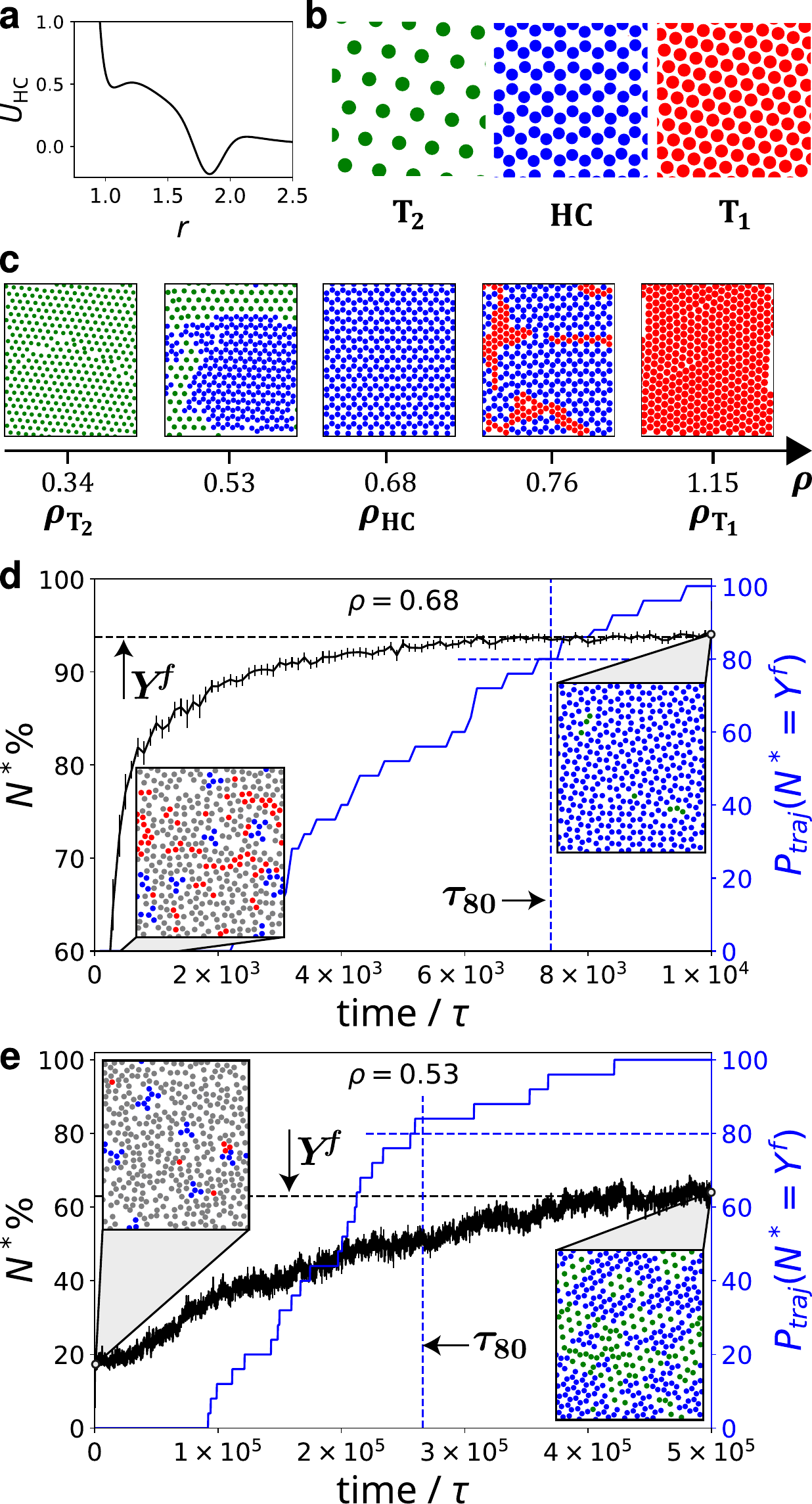}
    \caption{\textbf{Equilibrium and self-assembly behavior of the honeycomb NMO potential.}  \textbf{a.}~The NMO pair potential for HC self-assembly. \textbf{b.}~The sparse triangular (T$_2$), honeycomb (HC), and dense triangular (T$_1$) polymorphs. \textbf{c.}~Equilibrium configurations under the NMO potential at various densities. Particles are colored according to their local order: HC (blue), T$_1$ (red), and T$_2$ (green). \textbf{d--e.}~Ensemble-averaged self-assembly trajectories at $\rho=0.68$ (\textbf{d}) and $\rho=0.53$ (\textbf{e}).~The size of the largest contiguous cluster of HC-ordered particles, $N^*$ (black curve), is shown with error bars representing the standard error of the mean of $100$ trajectories.  The fraction of trajectories that have reached $Y^f_{\text{NMO}}$, $P_{\text{traj}}$ (blue curve), is used to determine the kinetic timescale $\tau_{80}$ (blue dashed line).  Insets show the initial and final configurations, in which disordered particles are colored gray.}
    \label{fig:den}
\end{figure}

The central features of pair potentials specified by \eqref{eq:pot} are two local minima at the nearest and second-nearest neighbor distances of the HC lattice, as well as two energy barriers between the local minima and to the right of the second minimum (\figref{fig:den}a).
These key features can be tuned by varying the design parameters, $\vec{a}$.
In Ref.~\cite{nmo2}, these parameters were chosen to optimize the thermodynamic stability of the HC lattice at a temperature near its melting point, such that the HC lattice is the minimum-free-energy microstructure over a range of number densities $\rho \in (0.675,0.7)$, where all number densities are reported in dimensionless units with respect to $\sigma^2$.
The resulting optimized potential is termed ``near-melting optimal'' (NMO, named after the optimization scheme in Ref.~\cite{nmo2}) potential throughout this work.
Subsequent studies of the complete phase diagram of the NMO potential~\cite{hc_phase} showed that below the melting temperature, $k_{\text{B}}T \simeq 0.2$, the HC lattice coexists with a dense triangular lattice (T$_1$) for $\rho \in (0.7,1.14)$ and with a sparse triangular lattice (T$_2$) for $\rho \in (0.34,0.675)$ (\figref{fig:den}b).
In the T$_1$ lattice, nearest-neighbor particles are separated by a distance corresponding roughly to the first local minimum of the pair potential, whereas in the T$_2$ lattice, nearest-neighbor particles are separated by a distance corresponding roughly to the second local minimum of the pair potential.

Equilibrium configurations of $N=384$ particles, obtained via molecular dynamics simulations (see \appref{app:lang}), are shown at various densities in \figref{fig:den}c.
To identify particles with a local HC environment, we use Steinhardt bond-order parameters~\cite{steinhardt} to assign all particles to locally HC (blue), T$_1$ (red), T$_2$ (green), or otherwise disordered (gray) configurations (see \appref{app:obj}).
At the temperature $k_{\text{B}}T = 0.1$, which we use throughout this work, we observe the T$_2$, HC, and T$_1$ phases at densities consistent with the results of Ref.~\cite{hc_phase}.

In this work, we are interested in self-assembly from an initially disordered phase.
We therefore study the assembly behavior of the NMO potential by first preparing equilibrium disordered configurations at a high temperature $(k_{\text{B}}T=10)$.
These initial configurations (see, for example, the left-side insets of \figref{fig:den}d--e) are then quenched to $k_{\text{B}}T = 0.1$ and evolved under Langevin dynamics (see \appref{app:lang}).
Here we report all timescales as dimensionless quantities relative to $\tau \equiv \sigma \sqrt{m/k_{\text{B}}T}$, where $m$ is the particle mass.
Representative results for the NMO potential are shown for systems prepared both at $\rho_{\text{HC}}$ (\figref{fig:den}d) and at a lower density of $\rho=0.53$ (\figref{fig:den}e), at which the HC and T$_2$ phases are in coexistence at equilibrium.

To measure the degree of assembly, we compute the largest contiguous cluster of particles that are in the target HC configuration, which we denote by $N^*$.
We define the \textit{finite-time assembly yield}, which serves as a proxy for the thermodynamic behavior of the assembly process, to be $Y^f \equiv \langle N^*(t_{\text{finite}}) \rangle$, where a fixed $t_{\text{finite}}$ is chosen empirically based on the approximate time at which $N^*$ plateaus under the NMO potential, and the average is taken over an ensemble of $100$ trajectories.
The choice of $t_{\text{finite}}$ does not significantly affect the results if chosen to be much longer than the initial timescale for HC cluster formation (see \textit{Supplementary Information}).
We note that $Y^f$ is a relevant thermodynamic-like quantity, even though it may not reflect the global free-energy minimum, since it characterizes the (potentially metastable yet extremely long-lived) microstructure that emerges from the assembly simulations.
To quantify the self-assembly kinetics, we define $P_{\text{traj}}$ to be the percentage of trajectories that have at some point crossed the finite-time assembly yield of the NMO potential, $Y^f_{\text{NMO}}$.
The logarithm of the \textit{self-assembly rate} is then characterized by $K \equiv \ln \tau^{-1}_{80}$, where $\tau_{80}$ is the earliest time at which $P_{\text{traj}}$ exceeds $80\%$.
For the example case of the HC NMO potential, \figref{fig:den}d--e shows both $N^*$ and $P_{\text{traj}}$ as a function of the assembly time, with $Y^f$ and $\tau_{80}$ indicated.
We emphasize that this approach for quantifying the assembly kinetics is robust even in scenarios where the NMO potential results in a low finite-time assembly yield, as is often the case when self-assembly is performed at low densities.
As a result, this definition consistently provides a large dynamic range over which kinetic optimization can be carried out.
Nonetheless, we stress that alternative reasonable definitions of the assembly kinetics, including, for example, different definitions of $P_{\text{traj}}$ and/or $K$, lead to consistent conclusions throughout this work (see \textit{Supplementary Information}).

\section{Multi-objective optimization using active learning}

\begin{figure}
\centering
\includegraphics[width=0.95\columnwidth]{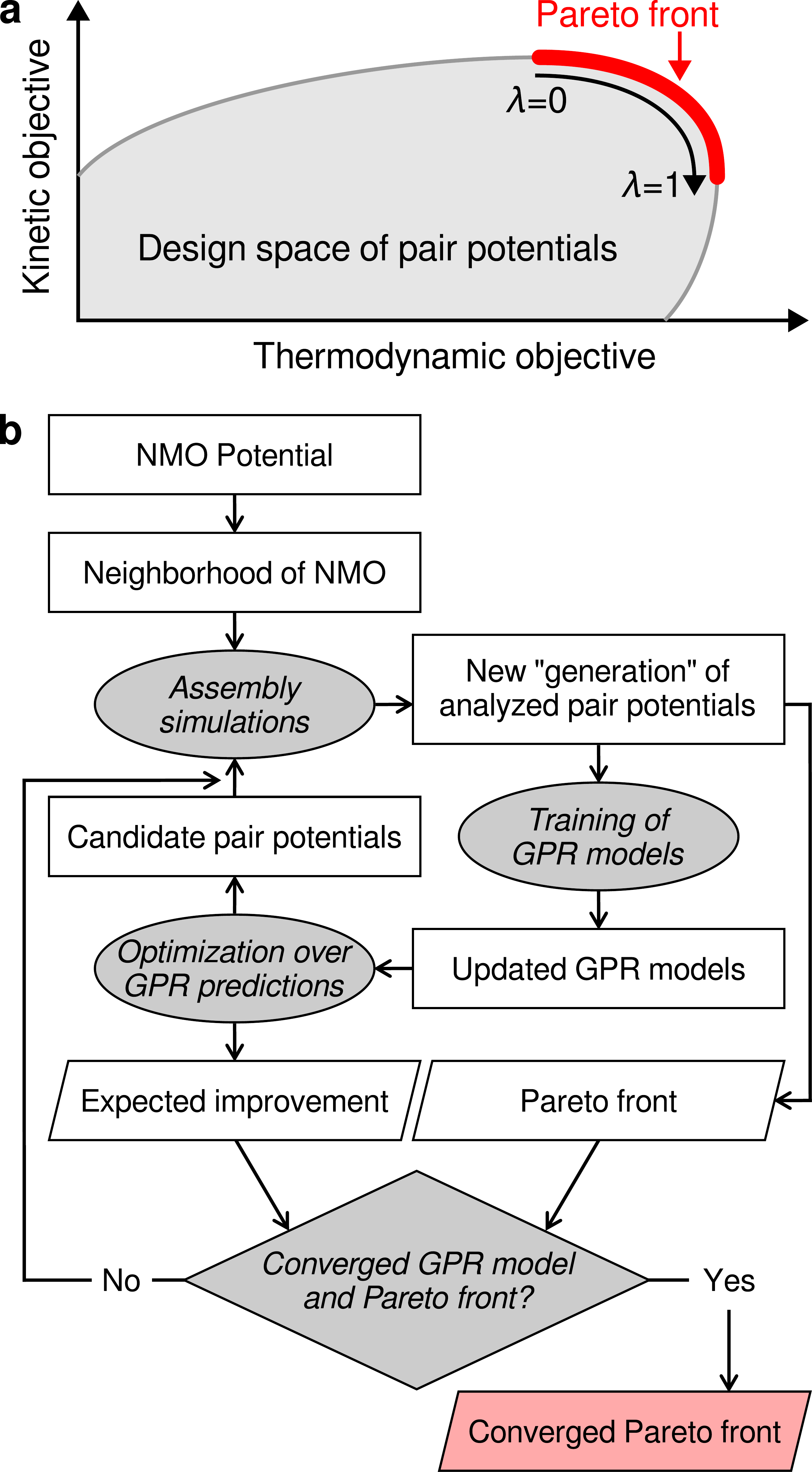}
\caption{\textbf{Identification of the thermodynamics/kinetics Pareto front via active learning.}  \textbf{a.}~Schematic of the Pareto front, which represents the pair potentials beyond which neither objective can be improved without sacrificing the other objective.  The variable $\lambda$ describes the front parametrically, such that the potential at $\lambda = 0$ is kinetically optimal, while the potential at $\lambda = 1$ is thermodynamically optimal.  \textbf{b.}~Schematic of the ML-guided iterative design algorithm for efficiently converging the Pareto front.}
\label{fig:gpr_schematic}
\end{figure}

Identifying pair potentials that simultaneously maximize the finite-time yield and self-assembly rate requires an exhaustive search of the design space.
In this section, we demonstrate that this goal can be achieved by performing multi-objective optimization via active learning.
To this end, we iteratively train Gaussian process regression (GPR) Bayesian surrogate models~\cite{gpr_ref} for the $Y^f$ and $K$ objectives, and then exploit these models to propose new design parameters for simulation.
By tracking the convergence of the GPR models, we are able to identify representative pair potentials along the Pareto front in the $Y^f$--$K$ plane (\figref{fig:gpr_schematic}a).

GPR surrogate models are useful for interpolating and extrapolating scalar quantities within a high-dimensional design space.
At points within the design space for which we have simulation data, we are able to specify the value and standard error of each objective precisely.
The GPR models then assume that both objectives vary smoothly within the design space according to an empirically chosen kernel (see \appref{app:gpr}), making it possible to predict each objective and the associated uncertainty elsewhere in the design space.
The choice of the kernel is crucial to the efficiency of this ML-guided optimization scheme, as it controls the predictive accuracy of the GPR models when trained on sparse data, but does not affect the Pareto front that is ultimately obtained from the converged models.

To initialize the GPR models, we first propose a set of design vectors $\{\vec{a}\}$ that are uniformly distributed near the parameters of the NMO potential (\figref{fig:gpr_schematic}b).
We then run assembly simulations starting from the disordered state (as described in \secref{sec2}) for each of the pair potentials specified by these design vectors, and compute $Y^f$ and $K$ to form the zeroth generation of our dataset.

Next, we train an independent GPR model for each of the objectives, $Y^f$ and $K$, using the zeroth generation of simulation data (see \appref{app:gpr} and Figure~S1 in \textit{Supplementary Information}).
We then use the trained GPR models to search for new potentials that maximize the \textit{expected improvement} of both objectives.
Expected improvement is a commonly used acquisition function in Bayesian optimization~\cite{exp_imp,ehvi1,ehvi2,ehvi3} that combines ``exploitation'' and ``exploration'' strategies by favoring regions of design space that are either predicted to improve the objective or, alternatively, to have a large uncertainty according to the surrogate model.
The expected improvement (EI) of an objective $Q$ is defined as
\begin{equation}
  \label{eq:EI}
    \text{EI}_Q \left( \vec{a} \right) \equiv s_Q(\vec{a})\left[ \gamma_Q(\vec{a}) \phi(\gamma_Q(\vec{a})) + \psi(\gamma_Q(\vec{a})\right],
\end{equation}
where the scaled improvement over the existing data is $\gamma_Q(\vec{a}) \equiv \left[ \mu_Q(\vec{a}) - Q(\vec{a}_{\text{best}}) \right]/s_Q(\vec{a})$; $Q(\vec{a}_{\text{best}})$ is the optimal value of $Q$ in the existing set of simulated design vectors; and $\mu_Q(\vec a)$ and $s_Q(\vec a)$ represent the mean and standard deviation (i.e., uncertainty), respectively, of the GPR prediction for $Q$ at $\vec a$.
The functions $\phi$ and $\psi$ are the probability density and cumulative distribution functions, respectively, of the normal distribution.
The first term in \eqref{eq:EI} promotes ``exploitation'' by maximizing improvement of the predicted mean relative to the current best solution, whereas the second term promotes ``exploration'' of regions of the design space in which the uncertainty is large.
We note that even when no further improvement to the mean is possible (i.e., when $\mu_Q(\vec{a}) = Q(\vec{a}_{\text{best}})$), $\text{EI}=\text{EI}_{\text{min}}$ remains nonzero since the second term is bounded from below by the finite uncertainty associated with the simulation data.

Since we wish to maximize the expected improvement of the Pareto front, we compute EI with respect to a weighted sum of the GPR predictions for the thermodynamic and kinetic objectives, $Q_\lambda \equiv \lambda Y^f + (1 - \lambda) K$, where $0\leq \lambda \leq 1$.
Computing $\text{EI}_{Q_\lambda}$ over the complete range of $\lambda$ values allows us to identify regions of the design space that are either predicted to advance any portion of the Pareto front to greater values of $Y^f$ and/or $K$ (\figref{fig:gpr_schematic}a), or are under-explored by the GPR model.
We note that this straightforward linear parameterization of the Pareto front (\figref{fig:gpr_schematic}a) does not introduce any assumptions regarding its shape.

\begin{figure*}
    \centering
    \includegraphics[width=0.95\textwidth]{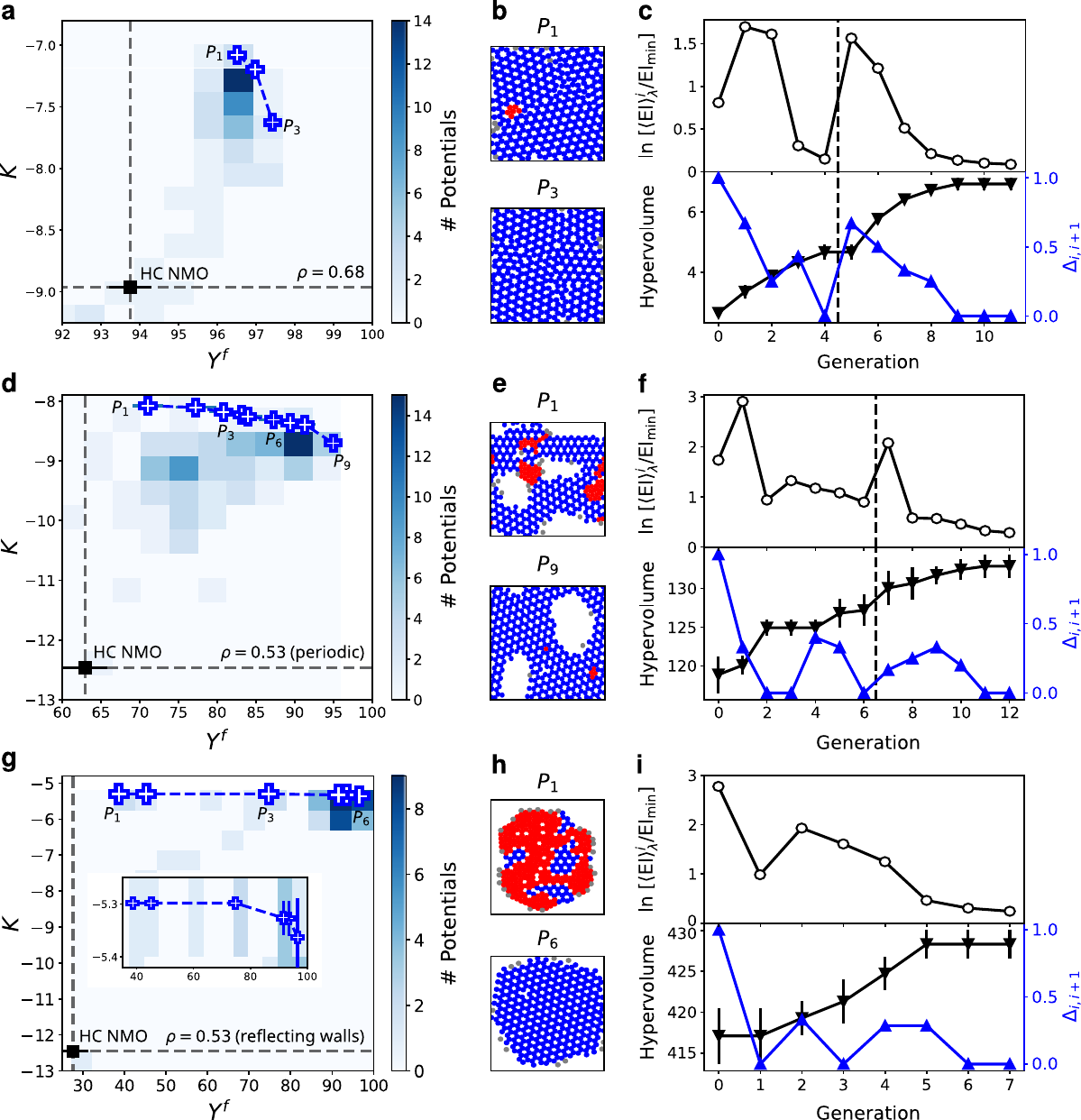}
    \caption{\textbf{Multi-objective optimization of honeycomb lattice pair potentials.}  \textbf{a.}~The converged Pareto front in the $Y^f$--$K$ plane at a density of $\rho = 0.68$ using periodic boundary conditions.  Pareto-optimal potentials, labeled from $P_1$ to $P_3$, are shown as crosses, while the NMO potential is indicated by a square.  The heat map (color bar) shows the distribution of the $112$ simulated candidate potentials in the $Y^f$--$K$ plane.  Because $K$ is defined on a logarithmic scale, the assembly of the most kinetically-optimal potential is nearly 10 times faster than that of the NMO potential.  \textbf{b.}~Representative long-time snapshots of the system under the most kinetically optimal ($P_1$) and thermodynamically optimal ($P_3$) potentials on the Pareto front. \textbf{c.}~Convergence of the average expected improvement, $\langle \text{EI} \rangle^i_\lambda$, relative to the lower bound established by the simulation uncertainty, $\text{EI}_{\text{min}}$ (black line and circles); the fraction of updated potentials on the Pareto front, $\Delta_{i,i+1}$ (blue line and triangles); and the hypervolume under the Pareto front (black line and inverted triangles) at each GPR generation.  Generations to the left of the dashed vertical line are obtained by global maximization of EI, while generations to the right are obtained by searching for local maxima of EI. \textbf{d--f.}~Converged Pareto front and distribution of the $120$ simulated candidate potentials, representative snapshots of system under Pareto-optimal potentials $P_1$ and $P_9$, and convergence analysis at a density of $\rho=0.53$ using periodic boundary conditions. \textbf{g--i.}~Converged Pareto front, representative snapshots under Pareto-optimal potentials $P_1$ and $P_6$, and convergence analysis at $\rho=0.53$ using reflecting wall boundary conditions. The GPR model in this scenario is initialized with the Pareto-optimal potentials found in \textbf{d}, and subsequent generations are obtained only by local maximization of EI.}
    \label{fig:gpr}
\end{figure*} 

Active learning proceeds by iteratively performing simulations for a generation of proposed design vectors, re-training the GPR models to include the latest generation of simulation results, and then selecting a new generation of design vectors that maximize the expected improvement according to the updated GPR models (\figref{fig:gpr_schematic}b).
This last step is accomplished by maximizing $\text{EI}_{Q_\lambda}$ using the trained GPR models, subject to prescribed bounds on the design vector, $\vec a$ (see \textit{Supplementary Information}).
We find that the Pareto front converges rapidly if we select new design vectors that globally maximize $\text{EI}_{Q_\lambda}$ in early iterations of ML-guided design, and then switch to proposing design vectors corresponding to local maxima of $\text{EI}_{Q_\lambda}$ in later iterations.
In practice, when applying this algorithm to the 4-dimensional design space of the HC pair potential family specified by \eqref{eq:pot}, we find that approximately 12 generations, consisting of $8$--$10$ proposed design vectors per generation in the global optimization stage and $10$--$12$ proposed design vectors per generation in the local optimization stage, are required to converge the $Y^f$--$K$ Pareto front (see, e.g., \figref{fig:gpr}a--c).
The number of simulations per generation scales with the number of $\lambda$ values used (see \appref{app:gpr}).
We emphasize that the efficiency of this approach is vastly superior to that of randomly sampling the design space, given that the probability of locating a Pareto-optimal potential by chance is $<0.1 \%$ in this example (Figure~S2 in \textit{Supplementary Information}).
Given that only $\sim 100$ simulated potentials are needed to identify $\sim 10$ Pareto-optimal potentials, ML-guided design is at least $100$ times more efficient than randomly sampling the design space of \eqref{eq:pot}.

We verify the convergence of the Pareto front by analyzing the prediction step of the ML-guided optimization algorithm and the changes to the Pareto front after every generation of simulations (see, e.g., \figref{fig:gpr}c).
The convergence of the GPR model itself is assessed by tracking the normalized average EI for the design parameters at the $i$th generation, $\langle \text{EI} \rangle^i_{\lambda}/ \text{EI}_{\text{min}}$, where $\text{EI}_{\text{min}}$ is approximated by averaging the uncertainty associated with the simulation data along the Pareto front.
This quantity transiently increases when switching from global to local maximization of EI, but otherwise tends to decrease with each generation.
Consistent with this decrease in the average EI, the fraction of potentials on the Pareto front that are updated at each generation tends to decrease with each generation.
Specifically, we compute $\Delta_{i,i+1} \equiv 1 - ( \text{PF}^{i+1|i}/\text{PF}^i )$, where $\text{PF}^{i+1|i}$ is the number of Pareto-optimal potentials in generation $i$ that remain in generation $i+1$, and $\text{PF}^i$ is the total number of Pareto-optimal potentials in generation $i$.
The fraction of updated potentials on the Pareto front is zero in the final generation.
We also show the convergence of the Pareto front by tracking the hypervolume under the Pareto front relative to an origin specified by the NMO potential (i.e., the area under the Pareto front within the integration bounds demarcated by black dashed lines in \figref{fig:gpr}a,d,g), which plateaus in the last $2$--$3$ generations.

\section{Self-assembly under Pareto-optimal potentials}

We now apply our active learning-based multi-objective optimization approach to a diverse set of test cases to demonstrate its generality.
We first consider self-assembly of the HC lattice (\figref{fig:gpr}).
Converged Pareto fronts under three different assembly conditions are shown in \figref{fig:gpr}a,d,g, where the Pareto front is approximated by a discrete set of Pareto-optimal pair potentials labeled $P_1$ through $P_m$.
Representative snapshots of the system at $t_{\text{finite}}$ under the most kinetically and thermodynamically optimal potentials on the Pareto front ($P_1$ and $P_m$, respectively) are shown in \figref{fig:gpr}b,e,h.
Convergence analyses are presented in \figref{fig:gpr}c,f,i.

In all cases, comparison with the HC NMO potential (indicated by a black square in \figref{fig:gpr}a,d,g) clearly shows that the Pareto-optimal potentials offer substantial improvements with respect to both the finite-time yield and the self-assembly rate.
Performing self-assembly at the higher density of $\rho = 0.68$ (\figref{fig:gpr}a) leads to a nearly 10-fold improvement in the self-assembly rate.
This observation is qualitatively consistent with the results of Ref.~\cite{goodrich}, which similarly found that the pair potential that maximizes an alternative measure of the self-assembly rate differs qualitatively from the free energy-optimized potential at this density.
More surprising, however, is the observation that the finite-time yield can also be improved relative to that of the HC NMO potential, which leads to kinetic arrest in a metastable state with lower $N^*$ (\figref{fig:den}d).
At the lower density of $\rho = 0.53$ (\figref{fig:gpr}d,g), we observe even greater kinetic and thermodynamic improvements of the Pareto-optimal potentials relative to the HC NMO potential, which leads to the assembly of kinetically arrested small clusters (\figref{fig:den}e).
These improvements relative to the NMO potential are similar regardless of whether periodic boundary conditions (\figref{fig:gpr}d) or reflecting wall boundary conditions (\figref{fig:gpr}g) are used, although we note that the absolute values of $K$ cannot be compared directly due to the different behavior of the HC NMO potential under periodic versus reflecting wall boundary conditions ($Y^f = 63\%$ versus $Y_f = 28\%$, respectively).

Although the qualitative features of the Pareto fronts shown in \figref{fig:gpr} reflect the physics of self-assembly, the precise shape of the front may be affected by the details of the objective definitions and constraints on the design space.
For example, Figure~S3 in \textit{Supplementary Information} shows that the Pareto front shifts vertically under a relaxed kinetic objective $K_{50} \equiv \ln \tau^{-1}_{50}$, since $\tau_{50}$ necessarily occurs before $\tau_{80}$.
Increasing $t_{\text{finite}}$ instead causes the kinetically optimal potentials to shift to lower values of $Y^f$ for reasons that are explained by our findings in the following section.
Similarly, the Pareto front expands slightly to lower values of $Y^f$ when we use an alternative kinetic definition based on a mean first passage time (Figure~S4 in \textit{Supplementary Information}).
Importantly, these definition-dependent effects are \textit{systematic}, so that the relative performance of each pair potential with respect to each objective, along with its position on (or off) the Pareto front, is unaffected to within the simulation uncertainty in the vast majority of cases.
We can also apply our method to a more generic family of pair potentials, in which we replace \eqref{eq:pot} with a cubic spline, $U_{\text{spline}}(r;\vec a)$, connecting $10$ equally spaced points $\{r_1, \ldots, r_{10}\}$; in this case, the design parameters $\vec a$ are the values of $U_{\text{spline}}$ at these 10 points.
The qualitative features of the Pareto front for HC self-assembly at $\rho=0.53$ are extremely similar between $U_{\text{HC}}$ and $U_{\text{spline}}$, suggesting that our results are not strongly affected by the functional form of \eqref{eq:pot} (Figure~S5 in \textit{Supplementary Information}).
However, the Pareto front for $U_{\text{spline}}$ is systematically shifted toward slightly suboptimal kinetic performance.
We interpret this shift as being a consequence of the finite number of points used in the definition of $U_{\text{spline}}$, which constrains the $r$-dependent variations in the forces between particles.
We nonetheless emphasize that the much higher dimension of the spline-potential design space highlights the efficiency of our active learning-based design approach.

\begin{figure}
    \centering
    \includegraphics[width=\columnwidth]{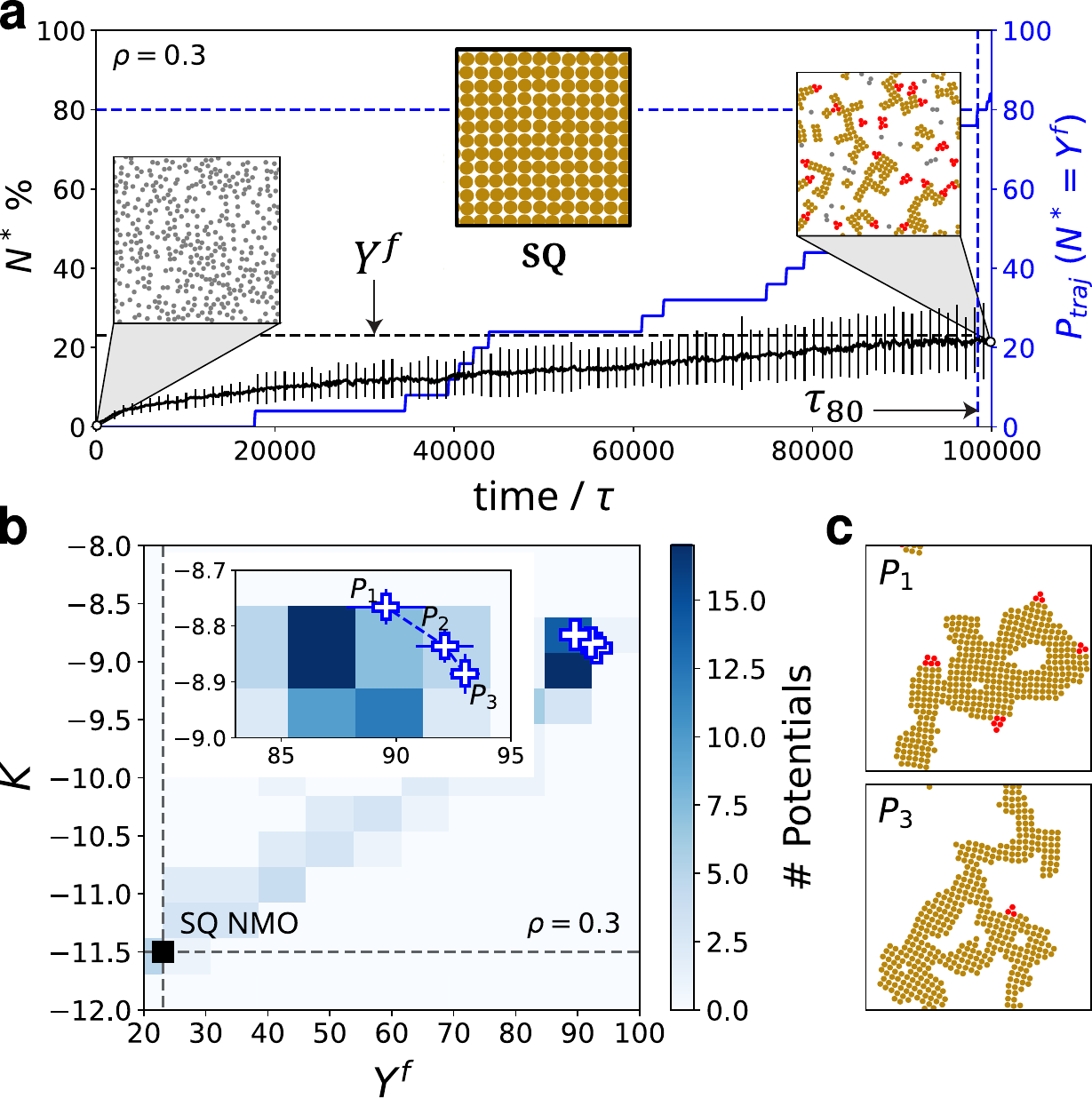}
    \caption{\textbf{Multi-objective optimization of square lattice pair potentials.} \textbf{a.}~Ensemble-averaged self-assembly trajectories under the SQ NMO potential, showing the size of the largest contiguous square cluster, $N^*$ (black curve), and the fraction of trajectories that have reached $Y^f_{\text{NMO}}$, $P_{\text{traj}}$ (blue curve).  Snapshots of the initial and final configurations, as well as the target square lattice, are shown as insets.  Particles are locally in square (gold), T$_1$ (red), or disordered (gray) configurations.  \textbf{b.}~The converged Pareto front.  \textbf{c.}~Representative snapshots of the system at $t_{\text{finite}}$ under the most kinetically optimal $(P_1)$ and thermodynamically optimal $(P_3)$ potentials.}
    \label{fig:square}
\end{figure}

We next consider self-assembly of the two-dimensional square (SQ) lattice, for which an NMO potential was identified in Ref.~\cite{nmo2}.
We work within the design space of the pair-potential family proposed in that work,
\begin{equation}
    U_{\text{SQ}}(r; \vec a) = \frac{1}{r^{12}} - \frac{2}{r^{6}} + a_0 e^{-a_1 \left( r-a_2 \right) ^2},
    \label{eq:sqpot}
\end{equation}
and follow an analogous protocol to that of the HC assembly simulations (see \appref{app:lang}).
Whereas the SQ NMO potential spontaneously assembles the square lattice at the density $\rho_{\text{SQ}}=1$ (Figure~S6 in \textit{Supplementary Information}), we test the efficacy of our ML-guided design approach by optimizing SQ self-assembly at a much lower density of $\rho=0.3$ using periodic boundary conditions.
Under these conditions, the SQ NMO potential leads to small, kinetically arrested SQ and T$_1$ clusters (\figref{fig:square}a).
By contrast, the Pareto-optimal potentials lead to considerably higher finite-time yields and faster self-assembly rates relative to the SQ NMO potential (\figref{fig:square}b,c); however, we note again that the absolute values of $K$ cannot be compared directly with those for HC assembly due to the lower finite-time yield of the SQ NMO potential ($Y^f = 23\%$).
The associated convergence analysis is presented in Figure~S7 in \textit{Supplementary Information}.

In summary, we find that Pareto-optimal pair potentials consistently out-perform the NMO potentials, both in terms of the finite-time yield and the self-assembly rate, when target lattices are assembled from an initially disordered mixture.
This observation applies across different target crystal polymorphs, densities, boundary conditions, and pair potential families.
In particular, the extent of improvement relative to the NMO potentials tends to be greater at lower densities, in which case the target lattice cannot fill the entire simulation box.
In all these test cases, our active learning-based multi-objective approach converges the Pareto front using surprisingly few simulations.
The design parameters of the NMO and Pareto-optimal potentials for HC and SQ self-assembly, along with the associated finite-time yields and self-assembly rates, are provided in Tables S1--S4 in \textit{Supplementary Information}.

\section{Multi-objective trade-offs and competing polymorphs}
\label{sec:tradeoff}

Whereas comparisons between Pareto-optimal and NMO potentials demonstrate the importance of considering dynamical processes as part of inverse design, comparisons among Pareto-optimal potentials reveal trade-offs between optimal self-assembly speeds and yields that are independent of any individual design algorithm.
We therefore analyze the Pareto-optimal potentials to understand whether optimizing for self-assembly speed alone will necessarily lead to a high finite-time yield, and vice versa, as well as the mechanistic origins of any trade-offs between these objectives.

Qualitative differences among the Pareto fronts for different target structures and initial densities reflect varying extents of a thermodynamics/kinetics trade-off for self-assembly.
If the endpoints of the Pareto front (i.e., at $\lambda = 0$ and $\lambda = 1$) are nearby in the $Y^f$--$K$ plane, then optimizing for either objective alone leads to similar results.
By contrast, if the Pareto front is extended, then the most kinetically and thermodynamically optimal potentials result in substantially different assembly trajectories.
We find that HC self-assembly at $\rho = 0.68$ (\figref{fig:gpr}a) and SQ self-assembly at $\rho = 0.3$ (\figref{fig:square}b) follow the former scenario, whereas HC self-assembly at $\rho = 0.53$ (\figref{fig:gpr}d,g) is representative of the latter.
These differences are also apparent from comparisons of the representative configurations at the two extremes of the Pareto front, as shown in \figref{fig:gpr}b,e,h and \figref{fig:square}c.

Nonetheless, in all cases we find at most a two-fold increase in the self-assembly rate between the $\lambda=0$ and $\lambda=1$ Pareto-optimal potentials for any target structure under any conditions.
As noted previously, this conclusion is independent of the precise kinetic definition used for computing the Pareto front (Figures~S3 and S4 in \textit{Supplementary Information}).
Thus, even when the endpoints of the Pareto front are far apart, optimizing for $Y^f$ alone offers substantial improvement over the NMO potential and leads to reasonably fast assembly kinetics as well.
We return to the consequences of these observations in \secref{sec:implications}.

\begin{figure*}[htb]
    \centering
    \includegraphics[width=\textwidth]{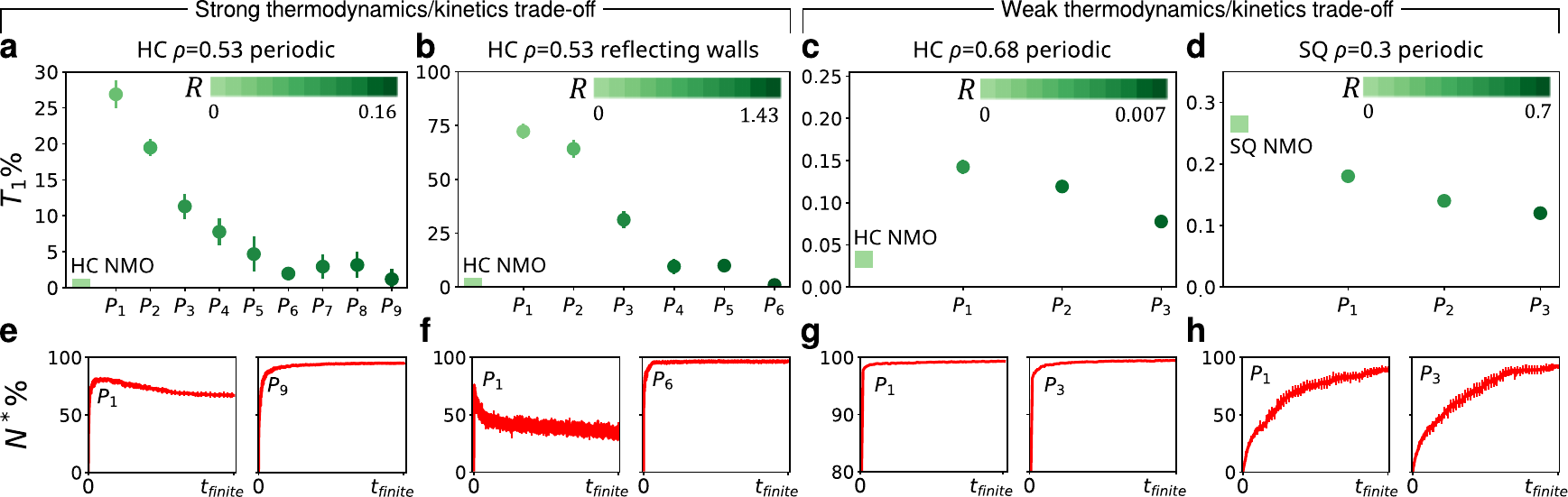}
    \caption{\textbf{Competition between target and T$_1$ polymorphs governs Pareto-optimal pair potentials.}  The fraction of T$_1$ defects at $t_{\text{finite}}$ under Pareto-optimal potentials for HC self-assembly at \textbf{a.}~$\rho=0.53$ with periodic boundary conditions, \textbf{b.}~$\rho=0.53$ with reflecting walls, and \textbf{c.}~$\rho=0.68$ with periodic boundary conditions, as well as SQ assembly at \textbf{d.}~$\rho=0.3$ with periodic boundary conditions.  Color bars show a scalar ``improvement metric'' $R \equiv \Delta Y^f \Delta K / Y^f_{\text{NMO}}|K_{\text{NMO}}|$, where $\Delta Y^f$ and $\Delta K$ indicate the improvement in each objective relative to the NMO potential.  $R$ increases from left to right across the Pareto front and is anti-correlated with the fraction of T$_1$ defects in all cases.  The HC NMO and SQ NMO potentials (which are not Pareto-optimal) are indicated by squares. \textbf{e--h.}~Ensemble-averaged assembly trajectories showing the largest cluster of the target polymorph, $N^*$, under the most kinetically and thermodynamically optimal potentials from the Pareto front corresponding to each scenario shown in \textbf{a--d}. Error bars represent the standard error of the mean of $100$ trajectories.}
    \label{fig:rho_tri}
\end{figure*}

The physical origins of this thermodynamics/kinetics trade-off can be understood by examining the formation of competing polymorphs during the assembly of a target lattice.
Motivated by the configurations shown in \figref{fig:gpr}b,e,h and \figref{fig:square}c, we compute the fraction of particles in local T$_1$ configurations at $t_{\text{finite}}$ for all Pareto-optimal potentials (\figref{fig:rho_tri}a--d).
The close-packed T$_1$ polymorph competes with both the HC and SQ lattices because the nearest-neighbor distance between the particles is the same.
We find that the fraction of T$_1$ defects at $t_{\text{finite}}$ decreases monotonically moving left-to-right across the Pareto front in all cases, suggesting that the assembly of this competing lattice governs the position of a Pareto-optimal potential along the front.
Moreover, the maximum fraction of T$_1$ defects correlates with the distance between the endpoints of the Pareto front, as can be seen by comparing \figref{fig:rho_tri}a,b and \figref{fig:rho_tri}c,d.
Finally, we note that the fraction of T$_1$ defects is non-zero for all Pareto-optimal potentials, whereas T$_1$ defects are sparse across all NMO potentials.
These long-time behaviors remain unchanged when $t_{\text{finite}}$ is doubled. 

This inverse relationship between assembly speed and the T$_1$ defect fraction along the Pareto front suggests that rapid assembly of a target lattice may lead to the formation of T$_1$ defects at later times.
This interpretation is corroborated in \figref{fig:rho_tri}e--h, where we observe two distinct classes of assembly trajectories.
For all near-thermodynamically optimal potentials, $N^*$ grows monotonically in time.
However, kinetically optimal potentials in scenarios where the endpoints of the Pareto front are far apart (specifically, \figref{fig:rho_tri}e,f) result in trajectories that are non-monotonic with respect to $N^*$, indicating that the rapid initial formation of the target structure is followed by later disassembly.
The formation of T$_1$ defects under kinetically optimal potentials is therefore even more pronounced at longer $t_{\text{finite}}$ (Figure~S3 in \textit{Supplementary Information}).
We emphasize that because these potentials are Pareto-optimal, this observation implies that maximizing the self-assembly rate \textit{necessarily} leads to T$_1$ defects at later times for these target structures and assembly conditions~\footnote{Competition between target-structure self-assembly and T$_1$ lattice formation, which governs the Pareto-optimal potentials, should not be confused with the observed coexistence between the HC lattice and the sparser T$_2$ lattice under the HC NMO potential at equilibrium. This distinction highlights the extent to which the phase behavior differs under the HC NMO and HC Pareto-optimal potentials}.

\begin{figure}
    \centering\includegraphics[width=\columnwidth]{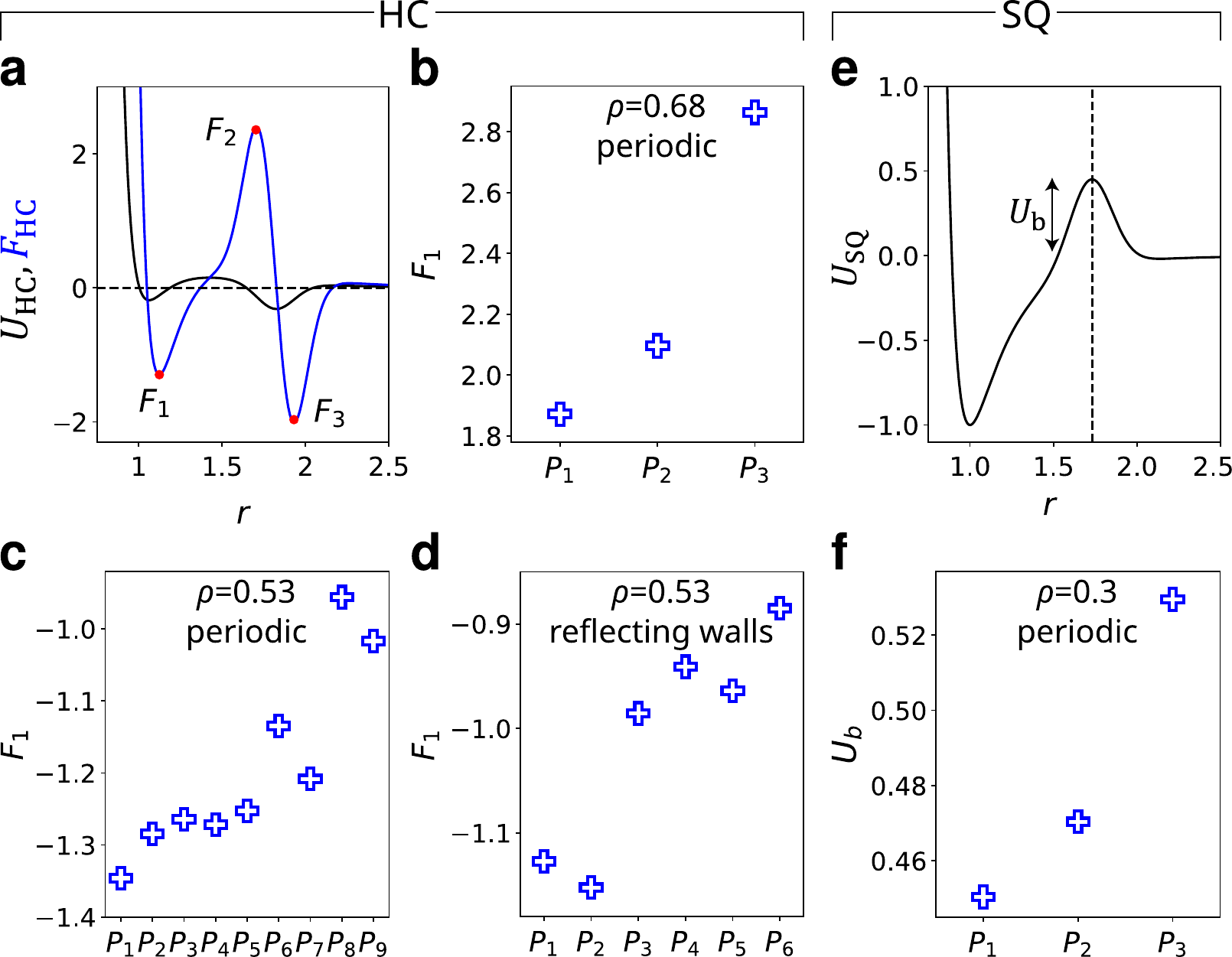}
    \caption{\textbf{The rank order of Pareto-optimal potentials correlates with key features of the potentials and forces.}  \textbf{a.}~Schematic of the force corresponding to the HC pair potential, \eqref{eq:pot}, highlighting the extrema $F_1$, $F_2$ and $F_3$. \textbf{b--d.}~$F_1$ values of Pareto-optimal potentials at $\rho=0.68$ with periodic boundary conditions, $\rho=0.53$ with periodic boundary conditions, and $\rho=0.53$ with reflecting wall boundary conditions. \textbf{e.}~Schematic of the SQ pair potential, \eqref{eq:sqpot}, highlighting the barrier height $U_{\text{b}}$.  The vertical dashed line shows the location of the second nearest neighbors in the T$_1$ polymorph. \textbf{f.}~$U_{\text{b}}$ values of Pareto-optimal potentials for SQ assembly at $\rho=0.3$ with periodic boundary conditions.}
    \label{fig:umap}
\end{figure}

Although the rank-ordering of the Pareto-optimal potentials does not correlate directly with specific design parameters (Figure~S8 in \textit{Supplementary Information}), the relative performance of these potentials can be interpreted in terms of features that drive T$_1$ defect formation.
Nonetheless, the details are target structure-dependent.
In the case of HC assembly, the maximum attractive force exerted by $U_{\text{HC}}$ close to the nearest-neighbor distance ($F_1$ in \figref{fig:umap}a) strongly correlates with the rank order of the Pareto-optimal potentials (\figref{fig:umap}b--d).
For all the assembly conditions that we consider, increasing the magnitude of this force increases the assembly rate but decreases the finite-time yield, likely because attractive forces at this distance favor both rapid HC assembly and, ultimately, conversion to the T$_1$ polymorph.
This finding is interesting because the dependence on forces as opposed to energetic features is a clear signature of optimization for dynamical quantities as opposed to free energies.
Pareto-optimal potentials obtained using either alternate kinetic definitions or the spline potential, $U_{\text{spline}}$, are similarly rank-ordered with respect to $F_1$ (Figure~S9 in \textit{Supplementary Information}). 

By contrast, the rank-ordering of the Pareto-optimal SQ potentials correlates with a feature of the pair potential, \eqref{eq:sqpot}, instead of a feature of the corresponding force.
This family of pair potentials features a barrier at the second nearest-neighbor distance of the T$_1$ polymorph (\figref{fig:umap}e).
Consequently, increasing the height of this barrier, $U_{\text{b}}$, tends to destabilize the T$_1$ polymorph.
We find that the most kinetically optimal potentials have a lower barrier height at this location, which promotes rapid self-assembly but likely increases T$_1$ polymorph formation at long times (\figref{fig:umap}f).
Thus, although the mechanistic details differ across target lattices, the rank ordering of the Pareto-optimal potentials appears to be consistently determined by the interplay between the target and competing polymorphs during self-assembly.

Taken together, these analyses highlight the crucial role of competing polymorphs in determining the extent of the thermodynamics/kinetics trade-off, as measured by the distance between the endpoints of the Pareto front, as well as the relative positions of the Pareto-optimal potentials along the front and the essential features of the pair potentials or the associated forces between particles.
Importantly, the identity of the competing polymorph (which is coincidentally the T$_1$ lattice in all cases we have studied here, but may be different in other self-assembly problems) and the existence of a thermodynamics/kinetics trade-off are not inputs to the active learning-based design algorithm, but rather emerge through optimization and \textit{post hoc} analysis.

\section{Implications for trajectory ensemble optimization}
\label{sec:implications}

The fundamental trade-offs revealed by the Pareto front also govern the behavior of optimization schemes that seek to identify optimal pair potentials without performing a global, multi-objective search.
In particular, gradient-based optimization schemes that utilize only short trajectories have been proposed as attractive alternatives, as they are amenable to computationally efficient automatic differentiation~\cite{goodrich} and path ensemble approaches~\cite{bolhuis2023}.
Therefore, in this section, we show how the shape of the Pareto front affects the outcomes of optimization schemes that only account for the self-assembly dynamics at short times.

To this end, we consider a simple gradient-descent algorithm for maximizing the number of particles in the largest HC cluster, $N^*$, at the end of a short trajectory duration, $t_{\text{f}}$.
Following the approach developed in Ref.~\cite{fim}, we focus on the probability distribution of system configurations at time $t_{\text{f}}$, $P(t;\vec{a})$, assuming fixed values of the design parameters $\vec a$.
This distribution can then be used to construct the Fisher Information Metric (FIM), $\bm{I} \equiv \langle \partial_{a_i} \ln P~ \partial_{a_j} \ln P \rangle$, where the angle brackets indicate an average over the entire configuration space.
The FIM describes fluctuations in the distribution $P(t;\vec{a})$ owing to changes in the design parameters $\vec{a}$, and is formally defined as the second derivative of the Kullback--Liebler divergence~\cite{nat_evol,nat_grad}.
The FIM therefore measures the effective distance between the distributions in terms of the design parameters.
Accounting for the FIM yields a modified gradient-descent scheme~\cite{fim} for maximizing $N^*$,
\begin{equation}
  \label{eq:fim}
  \vec{a}_{i+1} = \vec{a}_i - l_{\text{rate}} \times \bm{I}^{-1} \cdot \vec{\bm{C}},
\end{equation}
where $l_{\text{rate}}$ is an empirical learning rate.
Unlike conventional gradient descent in which $\vec{\bm{C}}\equiv \partial \langle N^*(t_{\text{f}}) \rangle / \partial \vec{a}$ would represent the gradient, in this approach $\bm{I}^{-1} \cdot \vec{\bm{C}}$ acts as the ``natural gradient'' of $\vec{a}$.
Importantly, by accounting for the fluctuations in the configuration space, the update rule given in \eqref{eq:fim} tends to be superior to a conventional gradient-descent procedure that directly follows gradients in the space of design parameters.

\begin{figure*}
    \centering\includegraphics[width=\textwidth]{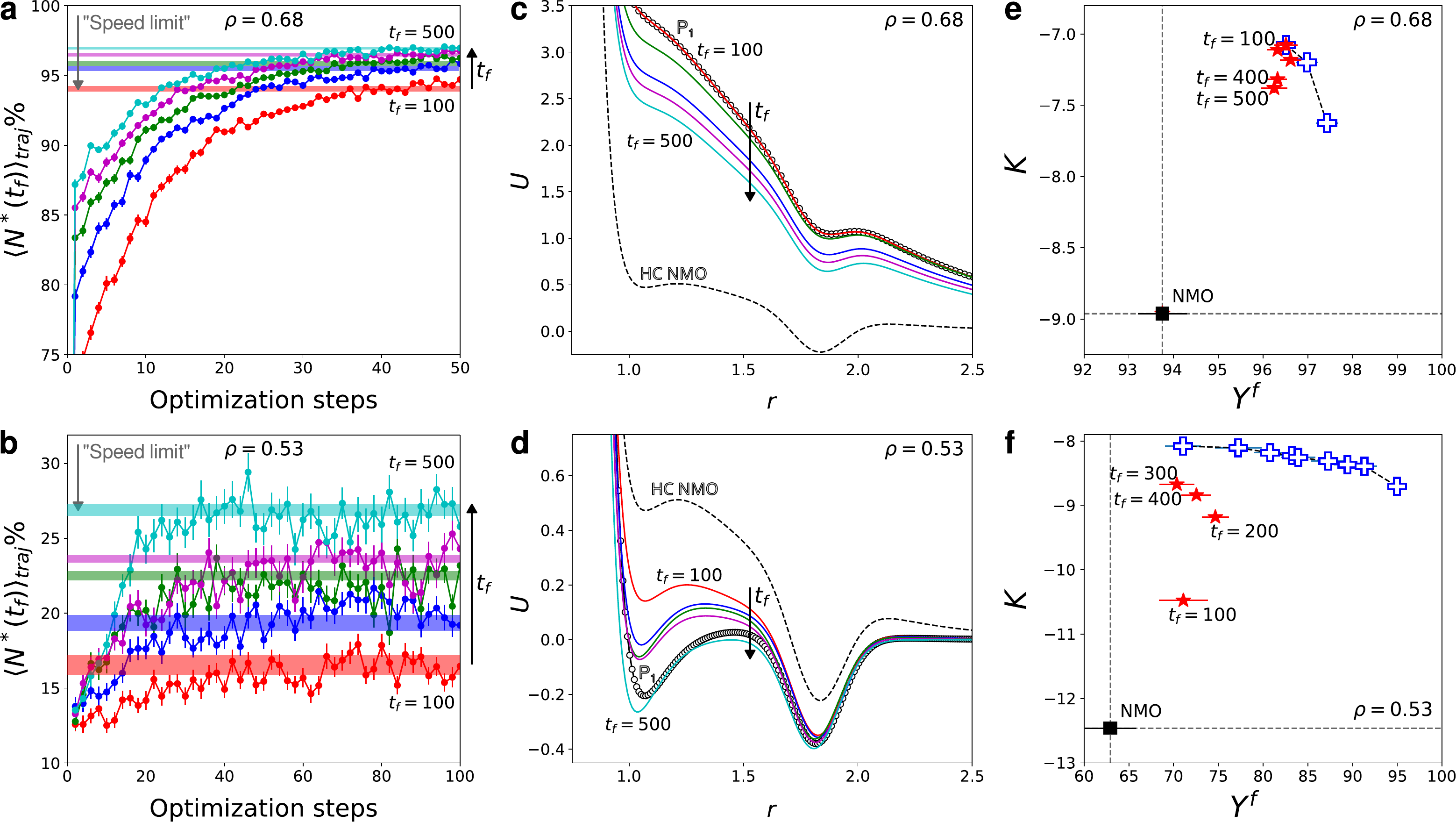}
    \caption{\textbf{HC self-assembly optimization using short trajectories.}  \textbf{a.}~Optimization at $\rho=0.68$ with periodic boundary conditions using trajectory durations of $t_{\text{f}}=100$ (red), $200$ (blue), $300$ (green), $400$ (magenta), and $500$ (cyan).  Points indicate the ensemble-averaged value of $N^*$ at $t_{\text{f}}$ after each optimization step, with error bars showing the standard error of the mean of $100$ trajectories.  For comparison, horizontal bars show $N^*(t_{\text{f}})$ for the most kinetically optimal potential on the Pareto front; the width of each bar indicates the standard error. \textbf{b.}~Optimization at $\rho=0.53$ with periodic boundary conditions using the same trajectory durations as in \textbf{a}. \textbf{c--d.}~Converged pair potentials at the conclusion of optimization at \textbf{c}.~$\rho=0.68$ and \textbf{d}.~$\rho=0.53$.  Each potential is colored as in panels \textbf{a} and \textbf{b}. For comparison, the HC NMO potential (dashed line) and the most kinetically optimal potential on the Pareto front ($P_1$, hollow black circles and line) are shown. \textbf{e--f.}~The performance of the converged pair potentials (red stars) in terms of the objectives $K$ and $Y^f$, with comparison to the Pareto-optimal potentials (crosses) and HC NMO potential (black square) shown in \figref{fig:gpr}a,d at \textbf{e}.~$\rho=0.68$ and \textbf{f}.~$\rho=0.53$.  Error bars indicate the standard error.  In \textbf{f}, the potential optimized at $t_{\text{f}}=500$ (not shown) leads to a long-time yield of $Y^f\approx50\%$.}
    \label{fig:optimization}
\end{figure*}

In what follows, we make the simplifying assumption that the system is in, or is sufficiently close to, a local equilibrium at time $t_{\text{f}}$.
This assumption greatly simplifies the computation of $\bm{I}$ and $\vec{\bm C}$ and turns out to be sufficiently accurate for converging optimal pair potentials in the specific examples that we consider here.
With this assumption of local equilibrium, the distribution over configuration space at time $t_{\text{f}}$ is given by a Boltzmann factor, $\hat{P} \propto \exp[-\beta U(r;\vec{a})]$, leading to the expressions
\begin{align}
  \label{eq:I}
  \tilde{\bm{I}} &\approx -\beta^2 \mathrm{Cov} \left[ \frac{\partial U(r;\vec{a})}{\partial \vec{a}}, \frac{\partial U(r;\vec{a})}{\partial \vec{a}} \right]_{\text{traj}}\!, \\
  \label{eq:C}
    \vec{\bm{C}} &\approx -\beta \mathrm{Cov} \left[ N^*(t_{\text{f}}),\frac{\partial U(r;\vec{a})}{\partial \vec{a}}\right]_{\text{traj}}\!,
\end{align}
where $\beta$ is the inverse temperature.
Both covariances are then evaluated over an ensemble of short trajectories.
Specifically, we launch $100$ independent trajectories starting from the disordered phase and evaluate \eqsref{eq:I} and (\ref{eq:C}) from the configurations obtained at time $t_{\text{f}}$.
Then, using a constant learning rate $l_{\text{rate}}=0.1$, we update the design parameters according to \eqref{eq:fim}.
(See Figure~S10 in \textit{Supplementary Information} for an analysis of the convergence with respect to the learning rate.)
Applying this algorithm to HC self-assembly, and starting from the HC NMO design parameters, we find that the potentials converge with respect to the ensemble-averaged value of $N^*$, $\langle N^* \rangle_{\text{traj}}$, within approximately 50 optimization steps at both $\rho = 0.68$ and $\rho = 0.53$ when using trajectory durations ranging from $t_{\text{f}} = 100$ to $t_{\text{f}} = 500$ (\figref{fig:optimization}a--b).
The resulting pair potentials are shown in \figref{fig:optimization}c--d.

For favorable choices of $t_{\text{f}}$, this short-trajectory optimization approach is capable of identifying pair potentials that approximate the kinetically optimal portion of the Pareto front (\figref{fig:optimization}e--f).
This approach performs best when the endpoints of the Pareto front are close to one another in the $Y^f$--$K$ plane, for example in the case of HC self-assembly at $\rho=0.68$.
However, even when the Pareto front is extended, as in the case of HC self-assembly at $\rho=0.53$, optimization using trajectory durations that are orders of magnitude shorter than $t_{\text{finite}}$ is surprisingly capable of improving both $K$ and $Y^f$ relative to the HC NMO potential.

To understand this behavior, we compare the average assembly at end of the short trajectories with a ``speed limit'' estimated from the full-length assembly trajectories, $\langle N^*(t_{\text{f}}) \rangle$, of the most kinetically optimal potential on the corresponding Pareto front (i.e., $P_1$ in \figref{fig:gpr}a,d).
In this way, we find that optimizing on the basis of $N^*(t_{\text{f}})$, i.e., \eqsref{eq:fim}--\ref{eq:C}, results in potentials that assemble the HC lattice to the same extent as the fastest assembling Pareto-optimal potential given the same trajectory duration (\figref{fig:optimization}a--b).
In fact, for HC self-assembly at $\rho=0.68$, optimization with respect to $N^*(t_{\text{f}}=100)$ essentially locates the Pareto-optimal potential $P_1$.
The qualitative features of this potential agree with the results of Ref.~\cite{goodrich}, which similarly optimized on the basis of short trajectories but used a different objective function and a different gradient-descent approach.
Choosing longer trajectory durations within this optimization scheme leads to progressively deeper first and second minima at both densities, with this effect being more pronounced for the first minimum (\figref{fig:optimization}c--d).
Thus, for HC self-assembly at $\rho=0.68$, this means that increasing the trajectory duration tends to bring the resulting potentials closer to the HC NMO potential, which we expect to be optimal at $t_{\text{f}} \rightarrow \infty$ since the HC polymorph is known to be the minimum-free-energy, space-filling lattice at this density (\figref{fig:optimization}c).
By contrast, at the lower density of $\rho=0.53$, longer trajectory durations lead the resulting potential farther from the HC NMO potential.

Nonetheless, since the short trajectories have no explicit information about the consequences of a particular potential at later times, this approach is unable to identify potentials with $Y^f$ values that are substantially greater than that of the left-most point on the Pareto front (\figref{fig:optimization}e--f).
This observation suggests a fundamental limitation of schemes that optimize by effectively seeking the speed limit of the most kinetically optimal potential on the Pareto front.
Of the examples shown in \figref{fig:optimization}, this problem is more severe at the lower density due to the greater thermodynamics/kinetics trade-off indicated by the elongated shape of the Pareto front (\figref{fig:optimization}f).
Moreover, since kinetically efficient potentials tend to eventually form a large fraction of T$_1$ defects (see \secref{sec:tradeoff}), increasing the short-trajectory duration beyond $t_{\text{f}} \approx 400$ actually leads to lower values of $Y^f$ compared to the NMO potential.
Choosing an appropriate value of the trajectory duration, $t_{\text{f}}$, is thus critical for locating the Pareto front using this optimization approach when it is applied to self-assembly problems with a significant thermodynamics/kinetics trade-off, and yet the criteria for selecting this parameter are generally unknown \textit{a priori}.
It is likely that this difficulty arises for short-trajectory-based optimization methods more generally.

\section{Discussion}

Practical algorithms for designing materials that form via self-assembly must optimize for both the final yield of a target structure and the time that it takes for the structure to form.
Our analysis demonstrates that the optimal solutions to prototypical self-assembly problems are generically governed by a thermodynamics/kinetics trade-off, but that the extent of this trade-off depends on the target structure and the assembly conditions.
Our central result is that this trade-off consistently arises due to a competition between alternative polymorphs, which allows us to explain key features of the thermodynamics/kinetics Pareto fronts and the optimized pair potentials that comprise them.
These conclusions emerge from our calculations and characterizations of Pareto-optimal pair potentials, which are enabled by an efficient active learning-based design algorithm that performs global optimization in a high-dimensional space of design parameters.
Importantly, this methodology can be straightforwardly applied to a wide variety of self-assembly problems with arbitrary design spaces, without needing to specify competing polymorphs \textit{a priori}.
Our approach therefore adds another useful tool to the growing list of inverse design methods that utilize ML techniques~\cite{ml_inv_design,truskett2020,dijkstra_ml_SA,truskett2016,goodrich,jhaveri2024discovering,Sulc2020,dijkstra2022,wang_inv,neuro_whitelam}.

Our results reveal that Pareto-optimal pair potentials are strongly influenced by the assembly conditions, and in general offer dramatic improvements in the self-assembly rate and yield compared to NMO potentials that were optimized on the basis of free-energy differences.
Key features of Pareto-optimal pair potentials can also be qualitatively different from those of the NMO potentials.
These comparisons emphasize that designing for thermodynamic equilibrium versus a finite-time yield, which may reflect an extremely long-lived metastable state, can be substantially different tasks.
Nonetheless, both approaches ultimately center on the avoidance of competing polymorphs, whether by construction in equilibrium methods such as Refs.~\cite{nmo1} and \cite{nmo2}, or organically due to the definitions of the optimization objectives in this work.
The distinction is that most equilibrium methods focus on a small number of configurational macrostates, but do not consider the possibility of other kinetically trapped configurations that arise during self-assembly and can reduce the yield at finite (but still extremely long) times.
Whereas the avoidance of trapped configurations can be specified via additional \textit{ad hoc} constraints in equilibrium approaches, such as Ref.~\cite{Sulc2020}, global optimization based on dynamical trajectories naturally accounts for kinetic traps when optimizing the objective(s).
Unexpected mechanistic details regarding the competition among polymorphs, such as our finding that properties of the force between particles rank order the Pareto-optimal HC potentials, can then emerge from completely unbiased analyses.
Of course, the catch is that optimization based on direct simulation can be computationally expensive.
Thus, successfully carrying out direct, global optimization of self-assembly trajectories for the types of canonical self-assembly problems that we consider requires an efficient ML-guided design algorithm, which we provide in this work.

Our results also supply useful guidance for the development of gradient-based inverse-design approaches that target kinetic self-assembly pathways.
Importantly, achieving a near-optimal finite-time yield by explicitly optimizing for the self-assembly rate alone only works when there is minimal competition among alternative polymorphs.
Optimization algorithms that rely solely on information from short trajectories, and thus seek the kinetically optimal portion of the Pareto front, are therefore likely to be of limited utility in scenarios with a substantial thermodynamics/kinetics trade-off.
We have provided an explicit example of this limitation by implementing a natural gradient descent optimization method for HC self-assembly using an ensemble of short trajectories.
While this gradient-descent algorithm successfully finds potentials close to the kinetically optimal region of the Pareto front, it cannot find potentials that optimize the yield on timescales that are orders of magnitude longer than the short trajectory durations if the endpoints of the Pareto front are distant from one another.
We expect that this limitation is likely to extend to other inverse-design methods based on short self-assembly trajectories.

In conclusion, we have shown that inverse design via active learning can be used to identify design parameters that maximize both the self-assembly rate and the finite-time yield, and also to provide physical insights into trade-offs inherent to a self-assembly process.
Our ML-guided optimization algorithm could easily be applied to more complex self-assembly processes with higher-dimensional design spaces, including multicomponent and polymeric systems.
Our approach could also be adapted to understand trade-offs in time-dependent self-assembly processes, potentially leading to an efficient strategy for performing protocol design~\cite{frenkel_td,neuro_whitelam,trubiano_td}.  
Finally, we emphasize that our approach could be applied to design problems that translate directly to experiments, such as the optimization of colloidal self-assembly experiments in which the interaction parameters are known with high accuracy~\cite{pine_dnacc}.
These avenues for future work promise to advance theoretical inverse-design techniques, bringing us closer to solving practical challenges faced in state-of-the-art experiments on complex self-assembly~\cite{jacobs2024assembly}.

\appendix

\section{Assembly simulations}
\label{app:lang}

To establish a prototypical yet realistic model of colloidal self-assembly, we investigate a 2-dimensional system of $N$ indistinguishable particles of mass $m$ and diameter $\sigma$ using Langevin dynamics.
The equation of motion is
\begin{equation}
    m \ddot{\bm{r}}(t) = - \nabla U(\bm{r}) - \gamma \dot{\bm{r}}(t) + \eta (t),
\end{equation}
where $\gamma$ the friction coefficient and $\eta(t)$ is a zero-mean Gaussian white noise with variance $\langle \eta (t) \eta (t') \rangle = 2\gamma k_{\text{B}}T\delta (t-t')$.
All simulations are performed at the temperature $k_{\text{B}}T=0.1$ with $m=\gamma=1$.
The potential energy function is defined in simulation units, so that all times have units of $\tau \equiv \sigma \sqrt{m/k_{\text{B}} T}$.
The initial disordered phase for assembly simulations is prepared by equilibrating the system at high temperature ($k_{\text{B}}T=10$) under the HC or SQ NMO potential.

HC self-assembly simulations are performed with $N=384$ particles.
At $\rho=0.68$, the periodic simulation box has dimensions $22.03\sigma \times 25.44\sigma$, which are chosen to match the HC unit cell at this density.
At $\rho=0.53$, the simulation box has dimensions $24.94\sigma \times 28.8\sigma$, and either periodic boundary conditions or reflecting wall boundary conditions are used.
SQ assembly simulations are performed with $N=400$ particles.
At $\rho = 0.3$, the simulation box has dimensions $36\sigma \times 37.037\sigma$ with periodic boundary conditions.

\section{Polymorph classification}
\label{app:obj}

The degree of self-assembly is determined by computing Steinhardt bond-order parameters for each particle $j$, $O^{j}_{\text{target}}= (1/n) \sum_{k=1}^{n} \exp(ni\theta_{jk})$, where $n$ is the number of neighboring particles within a cutoff distance slightly larger than the nearest-neighbor distance of the target lattice.
The angle $\theta_{jk}$ subtends the displacement vector between neighboring particles $j$ and $k$, $\vec r_k - \vec r_j$, and the horizontal axis.
For perfectly ordered particles, $|O^{j}_{\text{target}}|\equiv \sqrt{\operatorname{Re}[O^j_{\text{target}}]^2 + \operatorname{Im}[O^j_{\text{target}}]^2}=1$; however, due to thermal fluctuations, we consider particles to be ordered when $|O^j_{\text{target}}|>0.85$.
Finally, we define $N^*$ to be the largest contiguous cluster of locally ordered particles, and we compute this quantity using a depth-first-search algorithm.

The HC lattice has $n=3$ neighbors (within a cutoff distance $1.2\sigma$), T$_1$ and T$_2$ lattices have $n=6$ neighbors (within cutoff distances of $1.2\sigma$ and $2\sigma$, respectively), and the SQ lattice has $n=4$ neighbors (within a cutoff distance of $1.2\sigma$).
To distinguish between HC and T$_1$ lattices, which both have at least $3$ neighbors within the same cutoff distance, we impose the additional criteria that HC-ordered particles have $|O^j_{\text{HC}}>0.85|$ and \textit{exactly} $3$ neighbors.
Finally, particles with $|O^j_{\text{target}}<0.85|$ are marked as disordered, except when they occur at the boundary of a target-polymorph cluster, in which case they are assigned the same local order as the cluster.

\section{Gaussian process regression (GPR)}
\label{app:gpr}

We separately train two GPR models for the thermodynamic, $Y^f$, and kinetic, $K$, objectives.
Each GPR model utilizes the Mat{\'{e}}rn kernel~\cite{gpr_ref},
\begin{equation}
    k(d) = \frac{1}{\Gamma(\nu)2^{\nu-1}} \left( \frac{\sqrt{2\nu}}{l} d\right)^\nu K_\nu \left( \frac{\sqrt{2\nu}}{l} d\right),
\end{equation}
where $d$ is the Euclidean distance between two points in the parameter space, $K_\nu$ is a modified Bessel function, and $\Gamma$ is the Gamma function.
The hyperparameters $\nu$ and $l$, which control the smoothness and length scale of the kernel, respectively, are chosen separately for each GPR model by fitting the simulation data at each generation.
First, we randomly divide the data into equal-sized training and test sets, fit the GPR model with fixed hyperparameter values to the training set, and compute the Pearson correlation coefficient between the simulated and GPR-predicted data in the test set.
This procedure is repeated for $100$ random training/test splits of the data set to calculate an average Pearson correlation coefficient for the GPR predictions.
Next, we vary the kernel hyperparameters $\nu$ and $l$ to maximize the average Pearson correlation coefficient.
Once this correlation exceeds $0.9$, we consider the hyperparameters to be sufficiently well converged (see Fig.~S1 in \textit{Supplementary Information}).

We fit each GPR model using all the available simulation data and use the optimized kernels to make predictions for the next generation of simulations.
At each generation of our ML-guided design algorithm, we propose a new generation of design vectors by optimizing EI over the entire range of $\lambda$ at intervals of $\lambda = 0,0.1,0.2,\ldots,1$. 
In early iterations of ML-guided design, we globally maximize $\text{EI}_{Q_\lambda}$ using the simplicial homology global optimization (SHGO) algorithm~\cite{shgo} for each value of $\lambda$.
Then, once no new global maxima can be found using SHGO, we switch to searching for local maxima of $\text{EI}_{Q_\lambda}$ either using the Nelder--Mead~\cite{nm} algorithm or via gradient descent using the BFGS algorithm~\cite{bfgs}.
In this case, we perform many optimization runs for each value of $\lambda$, each time initializing the algorithm from randomly chosen points in the design space.
We then cluster the results of these Nelder--Mead/BFGS optimization runs to identify distinct local maxima of $\text{EI}_{Q_\lambda}$.
The design vectors corresponding to these maxima constitute the next generation of candidate pair potentials for simulation.
We note that the total number of maxima identified at each generation may be fewer than the number of $\lambda$ values, since the maxima for different $\lambda$ values may occur at nearly the same design parameters and thus end up clustered together.
When searching for local maxima, it is possible to obtain more distinct maxima in a generation than there distinct $\lambda$ values.

To facilitate the use of this methodology in future work, we provide codes for training and optimizing the kernels of the GPR models, as well as for proposing new generations of design vectors via either global or local EI maximization.
These codes are available at \texttt{https://github.com/wmjac/multiobj-self-assembly}.

%


\end{document}


\title{Supplementary Information for ``Multi-objective optimization for targeted self-assembly among competing polymorphs''}
\author{Sambarta Chatterjee}
\affiliation{Department of Chemistry, Princeton University, Princeton, NJ 08544, USA}
\author{William M. Jacobs}
\email{wjacobs@princeton.edu}
\affiliation{Department of Chemistry, Princeton University, Princeton, NJ 08544, USA}

\maketitle
\onecolumngrid

\setcounter{figure}{0}
\setcounter{subsection}{0}
\renewcommand{\theequation}{S\arabic{equation}}
\renewcommand{\thefigure}{S\arabic{figure}}
\renewcommand{\thetable}{S\arabic{table}}
\renewcommand{\thesubsection}{S\arabic{subsection}}

\section{GPR model training and convergence}

\begin{figure}[H]
    \centering
    \includegraphics[width=\textwidth]{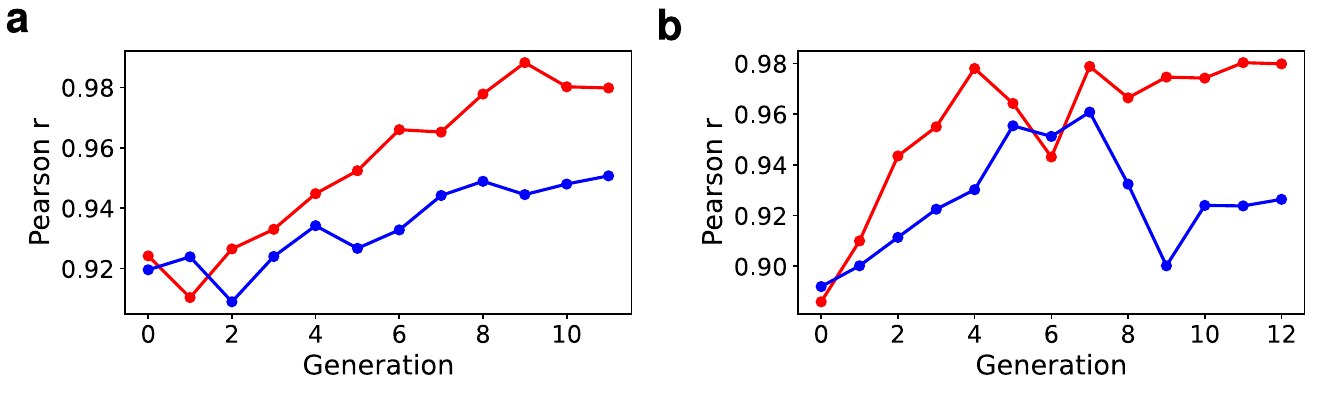}
    \caption{\textbf{GPR model consistency.} Pearson correlation coefficient (averaged over $100$ random training/test subsamples for each generation of data), for the $Y^f$ (red) and $K$ (blue) GPR models, for HC assembly at \textbf{a.}~$\rho=0.68$  and \textbf{b.}~$\rho=0.53$ with periodic boundary conditions.}
    \label{fig:training}
\end{figure}

The kernel for each of the independent thermodynamic and kinetic GPR models is fitted according to the procedure described in Appendix C of the main text. The Pearson correlation coefficient between randomly selected test data and predictions, averaged over $100$ such random subsamples, for the two GPR models for HC assembly at the densities $\rho=0.68$ and $\rho=0.53$, are shown in \figref{fig:training}. A high correlation coefficient verifies that the resulting GPR models, with optimized hyperparameters, describe the generation of data on which they were trained with high accuracy.

\section{NMO and Pareto-optimal potentials}

For HC self-assembly, we restrict the design space of the GPR search by imposing the bounds ${a_0 \in [5.0,6.5]}, {a_1 \in [1.0,100]}, {a_2 \in [1.0,5.0]}$, and ${a_3 \in [1.8,1.83]}$.
Potentials that lie outside these bounds do not lead to successful HC self-assembly, either because the first minimum is too attractive (forming the dense T$_1$ polymorph) or too repulsive (leading to a disordered dilute phase), or because the first or second minima are shifted to nearest-neighbor distances that are incompatible with the HC geometry assumed by our definition of $N^*$.
Potentials outside of these bounds are therefore not considered in the optimization process.
The design parameters of HC NMO and Pareto-optimal potentials for HC assembly at $\rho=0.68$ and $\rho=0.53$ (with periodic or reflecting wall boundary conditions), and their thermodynamic and kinetic objective values are summarized in Tables~S1--S3.

\begin{table}[H]
    \centering
    \begin{tabular}{||c c c c c c c||} 
        \hline
        Potential & $a_0$ & $a_1$ & $a_2$ & $a_3$ & $Y^f~(\%)$ & $K$ \\ [1ex] 
        \hline\hline
        HC NMO & $5.89$ & $17.9$ & $2.49$ & $1.823$ & $93.75$ & $-8.96$ \\
        $P_1$ & $5.99918$ & $18.1749$ & $1.36344$ & $1.81123$ & $96.51$ & $-7.0766$ \\
        $P_2$ & $5.7399$ & $16.415$ & $1.351$ & $1.83$ & $96.98$ & $-7.1978$ \\
        $P_3$ & $5.5$ & $17.088$ & $1.3$ & $1.83$ & $97.43$ & $-7.6245$ \\ [1.ex]
        \hline
    \end{tabular}
    \caption{HC assembly at $\rho=0.68$ with periodic boundary conditions, where the HC NMO potential equilibrates within $t_{\text{finite}} = 10^4 \tau$.}
    \label{table:1}
\end{table}

\begin{table}[H]
    \centering
    \begin{tabular}{||c c c c c c c||}
        \hline
        Potential & $a_0$ & $a_1$ & $a_2$ & $a_3$ & $Y^f~(\%)$ & $K$ \\ [1ex] 
        \hline\hline
        HC NMO & $5.89$ & $17.9$ & $2.49$ & $1.823$ & $63$ & $-12.462$ \\
        $P_1$ & $5.78242$ & $23.989$ & $3.5988$ & $1.82099$ & $71.06$ & $-8.078$ \\
        $P_2$ & $5.77569$ & $24.2511$ & $3.57752$ & $1.82108$ & $77.229$ & $-8.10455$\\
        $P_3$ & $5.76782$ & $24.4042$ & $3.5852$ & $1.82413$ & $80.789$ & $-8.181$\\
        $P_4$ & $5.7787$ & $24.8185$ & $3.5888$ & $1.82682$ & $83.187$ & $-8.225$\\
        $P_5$ & $5.77637$ & $24.999$ & $3.58756$ & $1.82682$ & $83.927$ & $-8.254$\\
        $P_6$ & $5.536$ & $15.159$ & $3.423$ & $1.816$ & $87.28$ & $-8.31$\\
        $P_7$ & $5.6138$ & $13.5705$ & $3.179$ & $1.823$ & $89.427$ & $-8.367$\\
        $P_8$ & $5.5$ & $15.499$ & $3.335$ & $1.823$ & $91.323$ & $-8.3925$\\
        $P_9$ & $5.5205$ & $16.6429$ & $3.4377$ & $1.829$ & $94.969$ & $-8.704$\\[1ex]
        \hline
    \end{tabular}
    \caption{HC assembly at $\rho=0.53$ with periodic boundary conditions, where the HC NMO potential becomes kinetically arrested within $t_{\text{finite}} = 5 \times 10^5 \tau$.}
    \label{table:2}
\end{table}

\begin{table}[H]
    \centering
    \begin{tabular}{||c c c c c c c||}
        \hline
         Potential & $a_0$ & $a_1$ & $a_2$ & $a_3$ & $Y^f~(\%)$ & $K$ \\ [1ex] 
         \hline\hline
         HC NMO & $5.89$ & $17.9$ & $2.49$ & $1.823$ & $27.55$ & $-12.45$ \\
         $P_1$ & $5.511$ & $13.4517$ & $3.3377$ & $1.8242$ & $38.594$ & $-5.298$\\
         $P_2$ & $5.60816$ & $16.06159$ & $3.337$ & $1.8238$ & $45.182$ & $-5.298$\\
         $P_3$ & $5.515$ & $15.66$ & $3.3456$ & $1.8256$ & $74.739$ & $-5.298$\\
         $P_4$ & $5.5059$ & $15.9566$ & $3.3404$ & $1.8239$ & $91.536$ & $-5.326$\\
         $P_5$ & $5.5111$ & $15.7418$ & $3.3375$ & $1.8242$ & $93.489$ & $-5.327$\\
         $P_6$ & $5.503$ & $16.7464$ & $3.3458$ & $1.82379$ & $96.588$ & $-5.363$\\
         \hline
    \end{tabular}
    \caption{HC assembly at $\rho=0.53$ with reflecting walls boundary conditions, using $t_{\text{finite}} = 5 \times 10^5 \tau$.}
    \label{table:3}
\end{table}

For SQ self-assembly, the design space was restricted for similar reasons to $a_0 \in [0.5,2]$, $a_1 \in [20,40]$, and ${a_2 \in [1.5,1.9]}$. The design parameters of the SQ NMO and Pareto-optimal potentials and their objective values are summarized in \tabref{table:4}.

\begin{table}[H]
    \centering
    \begin{tabular}{||c c c c c c||}
         \hline
         Potential & $a_0$ & $a_1$ & $a_2$ & $Y^f~(\%)$ & $K$ \\ [1ex]
         \hline\hline
          SQ NMO & $0.828$ & $26.5$ & $1.79$ & $23.11$ & $-11.497$ \\
         $P_1$ & $0.5243$ & $31.9812$ & $1.7234$ & $89.56$ & $-8.766$ \\
         $P_2$ & $0.5428$ & $31.3519$ & $1.7296$ & $92.13$ & $-8.836$ \\
         $P_3$ & $0.5859$ & $35.3879$ & $1.8061$ & $93$ & $-8.885$ \\
         \hline
         \end{tabular}
    \caption{SQ assembly at $\rho=0.3$, where the SQ NMO potential equilibrates within $t_{\text{finite}}=10^5 \tau$.}
    \label{table:4}
\end{table}

\section{Efficiency of ML-guided optimization compared to random sampling of the design space}
\begin{figure}[H]
    \centering
    \includegraphics[width=0.5\textwidth]{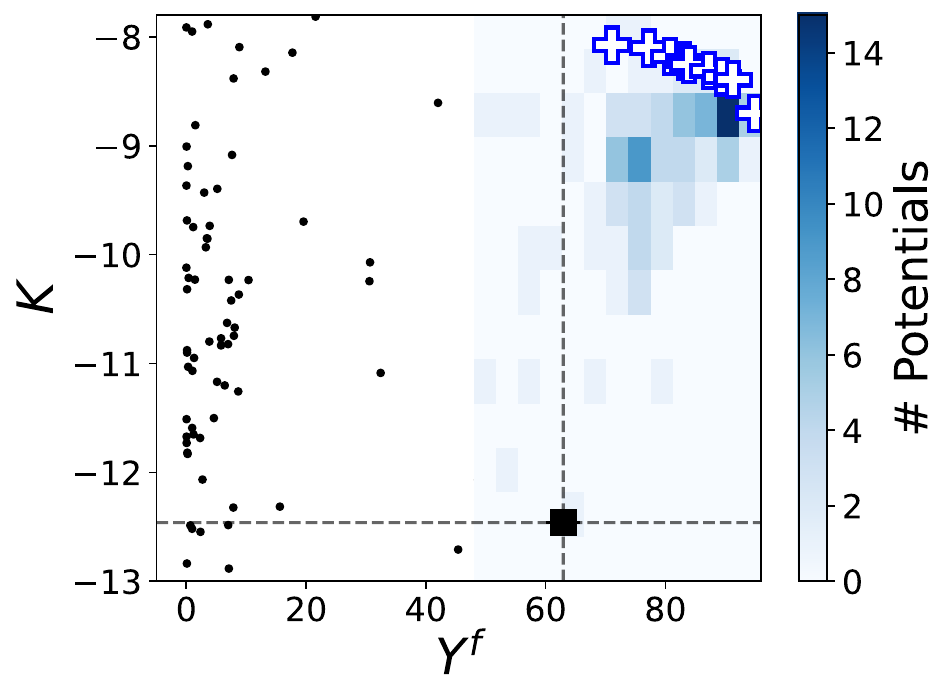}
    \caption{\textbf{Comparison of ML-guided design with random sampling.} Randomly sampled pair potentials (black dots) have less than a $0.1\%$ chance of locating the Pareto front (blue crosses) for HC assembly at $\rho=0.53$ with periodic boundary conditions.}
    \label{fig:random}
\end{figure}

\figref{fig:random} shows that a brute-force random sampling attempt to find the Pareto front is incredibly inefficient.
Given the HC design-space bounds provided in the previous section, and the fact that the Pareto-optimal potentials for HC assembly at $\rho=0.53$ using periodic boundary conditions lie within a region ${a^P_0 \in [5.5,5.782]}, {a^P_1 \in [13.57,24.99]}, {a^P_2 \in [3.179,3.599]}, {a^P_3 \in [1.816,1.829]}$, the probability of finding a Pareto optimal potential by random sampling is at best $P_{\text{random}}=\Pi_i \Delta a^P_i / \Pi_i \Delta a_i = 0.098 \%$.
By contrast, we show in the main text that our ML-guided design approach requires $\sim 100$ potentials to locate the Pareto front under these conditions.

\section{Pareto-front sensitivity analysis}
\begin{figure}[H]
    \centering
    \includegraphics[width=0.5\textwidth]{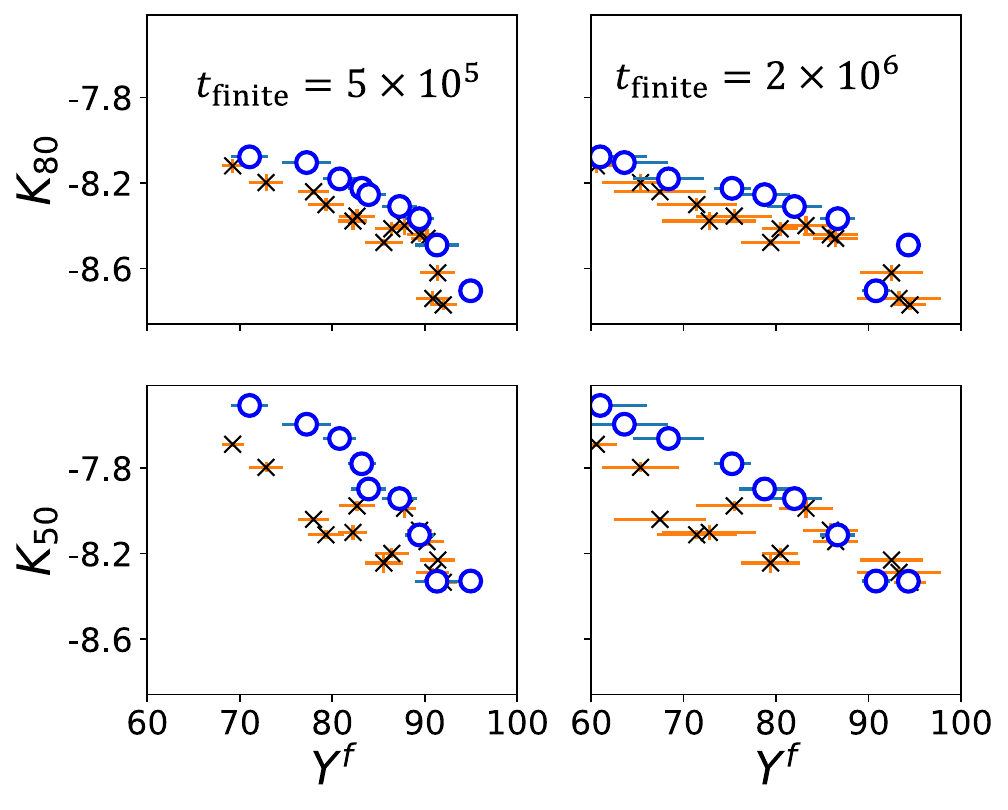}
    \caption{\textbf{Pareto front sensitivity analysis for HC assembly at $\rho=0.53$ with periodic boundary conditions}. Blue circles indicate the Pareto-optimal potentials reported in the main text. Black crosses indicate sub-optimal potentials, which may potentially become Pareto-optimal under different objective definitions.  Horizontal lines indicate the uncertainty in $Y^f$.}
    \label{fig:sensitivity}
\end{figure}

The robustness of the HC, $\rho=0.53$ Pareto front with respect to the proposed thermodynamic and kinetic objectives is confirmed by varying the simulation duration between $t_{\text{finite}} = 5\times 10^5 \tau$ and $2\times 10^6 \tau$, and the kinetic definition between $K_{50}\equiv \ln \tau^{-1}_{50}$ and $K_{80}\equiv \ln \tau^{-1}_{80}$, where the subscript indicates that either $50\%$ or $80\%$ of the assembly trajectories exceed $Y^f_{\text{NMO}}$, respectively.
\figref{fig:sensitivity} shows the Pareto front potentials reported in the main text, as well as a neighborhood of sub-optimal potentials, under these alternate definitions.
The kinetically optimal potentials perform worse in terms of $Y^f$ at longer $t_{\text{finite}}$ owing to greater T$_1$ defect formation.
By contrast, the more lenient kinetic objective systematically shifts the Pareto front to larger values in the vertical direction.

\begin{figure}[H]
    \centering
    \includegraphics[width=0.9\textwidth]{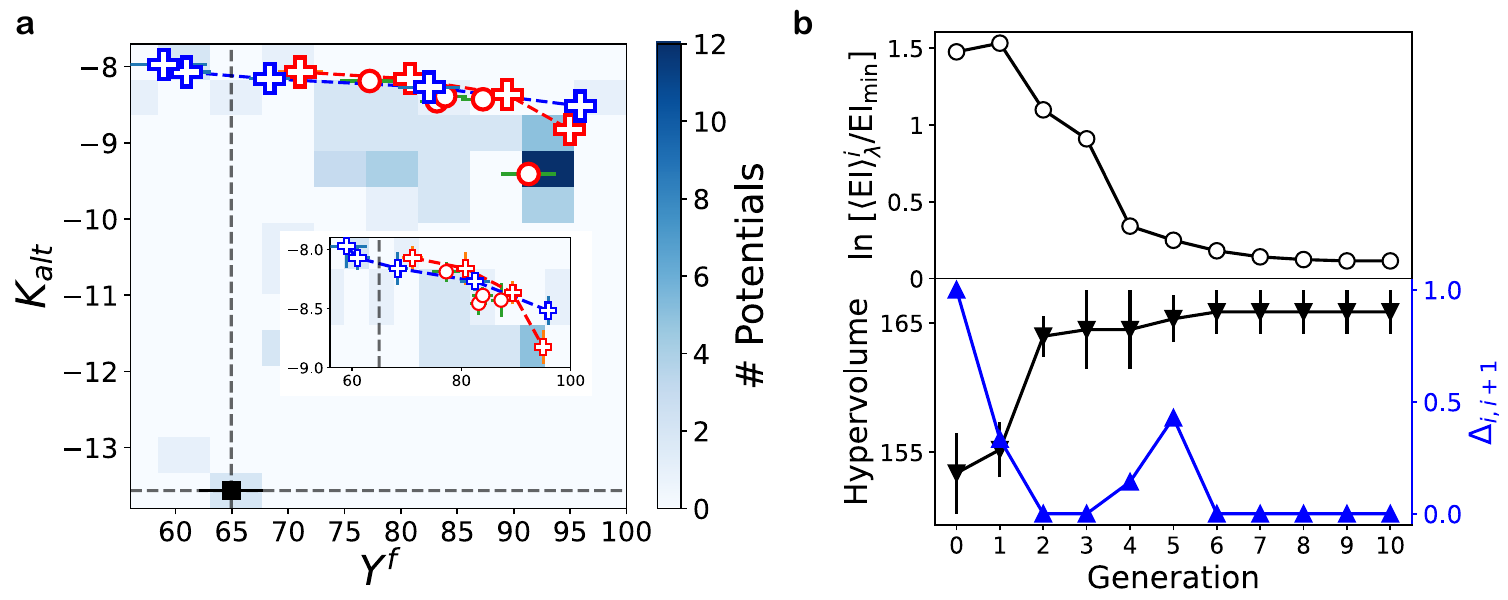}
    \caption{\textbf{Pareto front sensitivity under an alternate kinetic definition}. \textbf{a.}~Blue crosses indicate the converged Pareto front obtained using the alternate kinetic objective definition, $K_{\text{alt}}$, for HC self-assembly at $\rho = 0.53$ with periodic boundary conditions.  Red crosses and circles indicate the performance of the Pareto-optimal potentials presented in the main text under this alternate definition.  Red crosses indicate potentials that remain Pareto-optimal under the alternate kinetic definition, whereas red circles indicate potentials that are no longer Pareto-optimal under the new kinetic definition.  \textbf{b.}~The associated convergence analysis of the normalized average expected improvement (black line and circles), the hypervolume under the Pareto front (black line and inverted triangles), and the fraction of updated potentials on the Pareto front (blue line and triangles).}
    \label{fig:navg}
\end{figure}

We also consider an alternate kinetic definition, $K_{\text{alt}} \equiv \ln \tau_{Y^f_{\text{NMO}}}^{-1}$, where $\tau_{Y^f_{\text{NMO}}}$ is the time at which $\langle N^*\rangle$ crosses the NMO finite-time yield.
The uncertainty associated with this definition is $\Delta K=\Delta t_W / \tau_{Y^f_{\text{NMO}}}\! \sqrt{N}$. Here $\Delta t_W$ is defined as the time window $t_f - t_i$, where $t_i$ is the first time when $\langle N^* \rangle + \sigma_{N^*}/2 \geq Y^f_{\text{NMO}}$, $t_f$ is the first time at which $\langle N^* \rangle - \sigma_{N^*}/2 > Y^f_{\text{NMO}}$, and $\sigma_{N^*}$ is the standard deviation of $N^*$ across the 100 independent trajectories. 
\figref{fig:navg} shows a comparison between converged Pareto fronts using the original kinetic definition and this alternate kinetic definition.

Finally, to explore the sensitivity of the Pareto front with respect to the assumed family of pair potentials, we consider a more general functional form than that specified by Eq.~(1) in the main text.
To this end, we define a set of $12$ equally spaced $\{r_i\}$ values over the range $[r_{\text{min}},r_{\text{max}}]$, which we use to define a piecewise-continuous potential in the following manner.
We note that the Pareto-optimal potentials are fairly indistinguishable for $r\leq r_{\text{min}} = 1.0$, so we choose to use the value of the potential $P_8$ in \tabref{table:2} evaluated at $r_{\text{min}}$, $U^{(\text{P}_8)}(r_{\text{min}}) = 0.05$, to define the first point of the spline.
Similarly, we choose $r_{\text{max}} = 2.1$ based on the typical distance at which the Pareto-optimal potentials reach $U^{\text{P}_k}(r_{\text{max}}) =0$.
A cubic spline, $c(r;\{U_i\})$, is then defined by the points $\{ (1,0.05), (1.1, U_1), (1.2, U_2), \ldots, (2.0,U_{10}), (2.1,0)\}$, where the vector $\{U_1, \ldots, U_{10}\}$ represents the design parameters.
We use `not-a-knot' boundary conditions for $c(r;\{U_i\})$, which ensure that the third derivative of the polynomial is continuous at the end points.
Altogether, the spine potential is
\begin{equation}
  \label{eq:spline}
    U_{\text{Spline}} =
    \begin{cases}
        U^{(\text{P}_8)}(r), & r\le r_{\text{min}} \\
        c(r;\{U_i\}), & r_{\text{min}}<r<r_{\text{max}} \\
        0, & r \geq r_{\text{max}}.
    \end{cases}
\end{equation}

\begin{figure}[H]
  \centering
  \includegraphics[width=\textwidth]{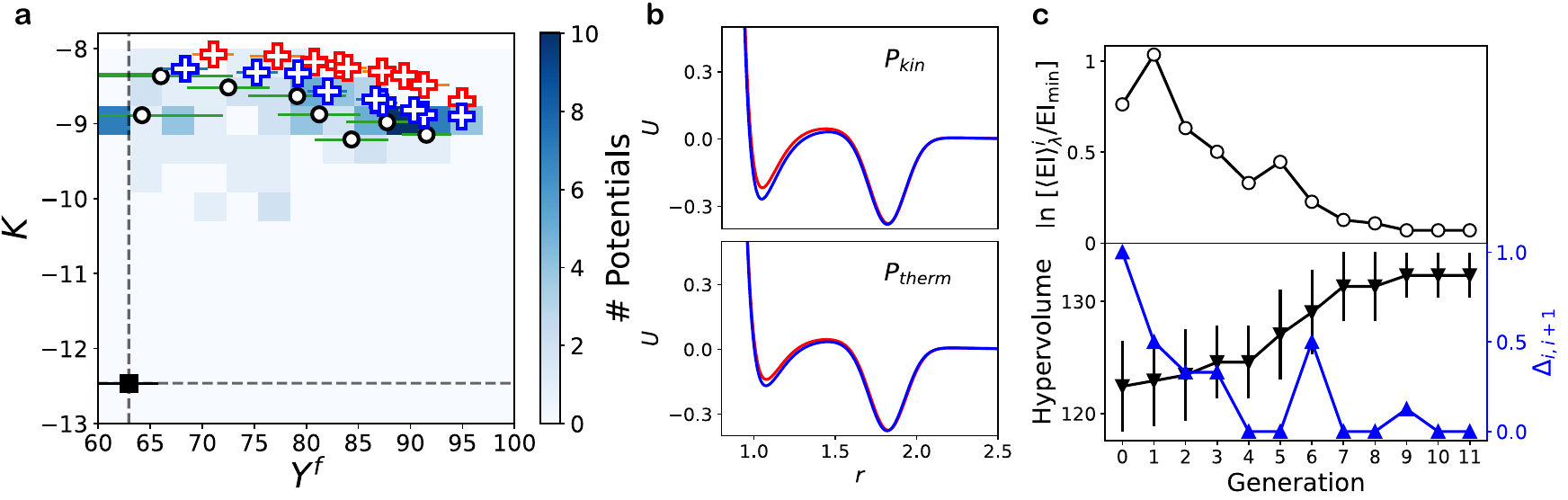}
  \caption{\textbf{Pareto front corresponding to spline-fitted potentials.} \textbf{a.}~Blue crosses indicate Pareto-optimal potentials found by ML-guided inverse design using the family of pair potentials specified by \eqref{eq:spline}. Red crosses indicate the original Pareto front for HC self-assembly at $\rho=0.53$ with periodic boundary conditions. Black circles indicate the initial fitted spline potentials, which form the zeroth generation for these GPR models. \textbf{b.}~The most kinetically and thermodynamically optimal potentials. Red lines represent the original Pareto-optimal potentials shown in the main text (red crosses in \textbf{a}). Blue lines represent the Pareto-optimal potentials found using the spline family of pair potentials (blue crosses in \textbf{a}). \textbf{c.}~Convergence analysis showing normalized average EI (black line and circles), the hypervolume under the Pareto front (black line and inverted triangles), and the fraction of updated potentials on the Pareto front (blue line and triangles).  See also \figref{fig:force} for further analysis.}
  \label{fig:spline}
\end{figure}

We seed the initial generation of the GPR models by fitting \eqref{eq:spline} to the Pareto optimal potentials shown in \tabref{table:2}, $U^{\text{P}_k}(r)$ for $k = 1, \ldots, 9$, to find the design parameters that minimize
\begin{equation}
  \Delta_{\text{fit}} \equiv \int_{r_{\text{min}}}^{r_{\text{max}}} dr\, r^2 \left[U^{\text{P}_k}(r) - U_{\text{spline}}(r)\right]^2.
\end{equation}
This approach ensures a least-squares fit over the range of distances where the potential is determined by the design parameters.
Thus, the zeroth generation of the GPR models consists of $9$ sets of $10$-dimensional design parameters, each corresponding to the best fit to a Pareto-optimal potential shown in \tabref{table:2}.
We then perform self-assembly simulations using these zeroth-generation potentials, and we proceed with the ML-guided design algorithm described in the main text.
\figref{fig:spline} shows the resulting converged Pareto front under this family of spline-fit potentials, \eqref{eq:spline}.

\section{Self-assembly and ML-guided optimization of the square lattice}

\begin{figure}[H]
    \centering
    \includegraphics[width=0.75\textwidth]{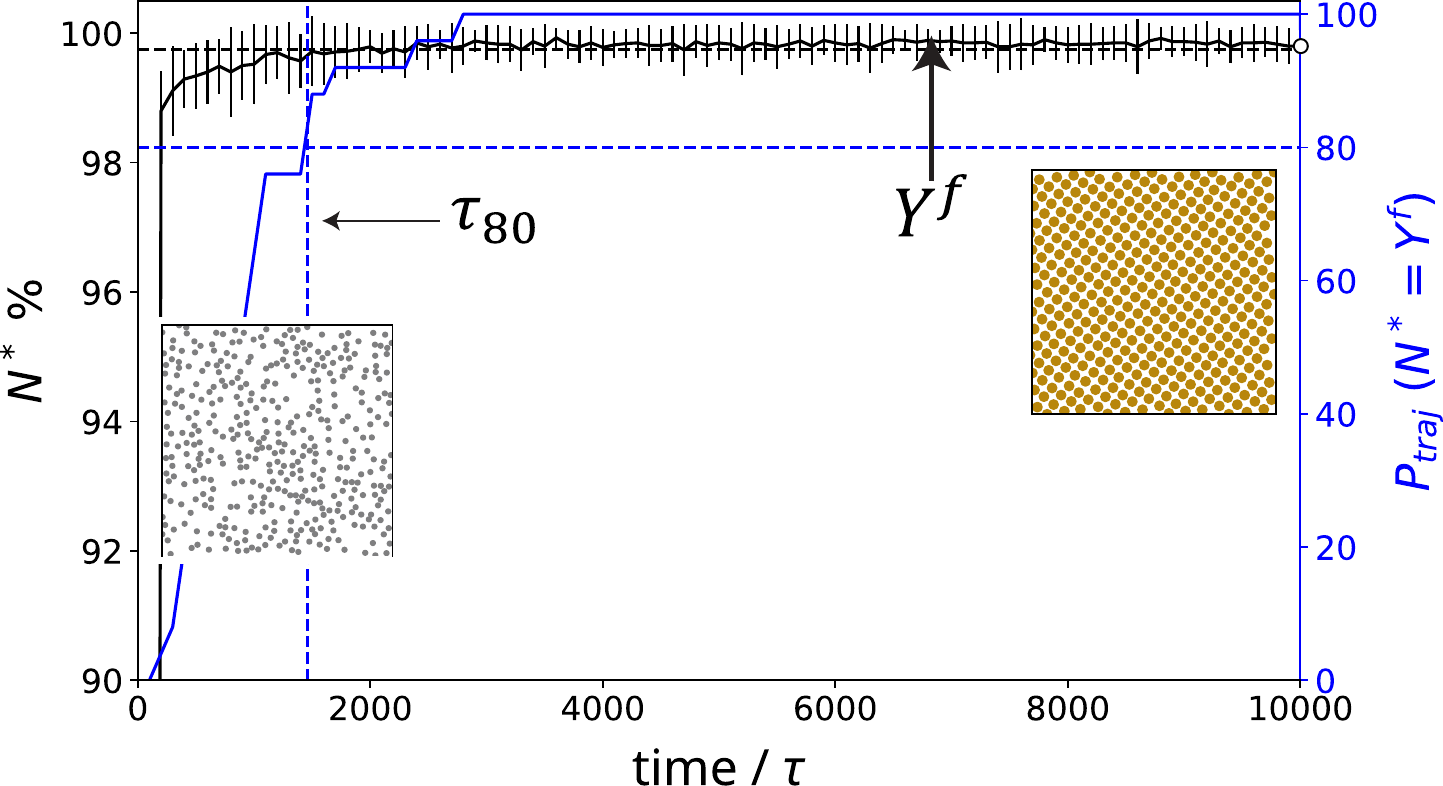}
    \caption{\textbf{Self-assembly under the SQ NMO potential at the density $\rho_{\text{SQ}}=1$}. Ensemble-averaged self-assembly trajectories showing the size of the largest contiguous cluster, $N^*$ (black curve), and the fraction of trajectories that have reached $Y^f_{\text{NMO}}=99.75\%$. Insets show representative snapshots of the initial and final configurations.}
    \label{fig:sq-nmo}
\end{figure}

\begin{figure}[H]
    \centering
    \includegraphics[width=0.85\textwidth]{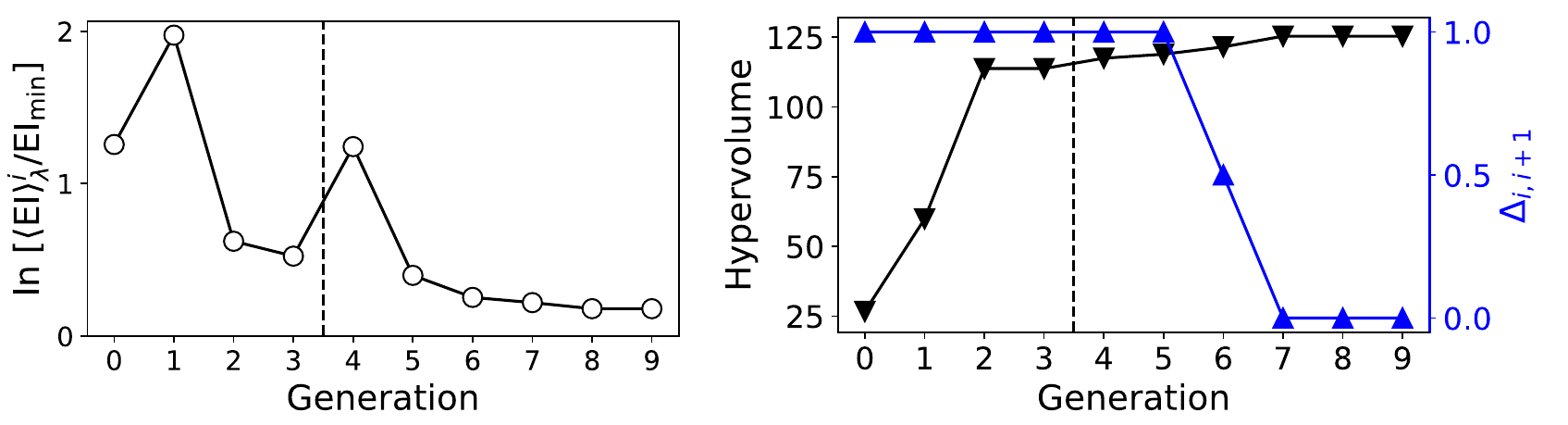}
    \caption{\textbf{Convergence of ML-guided design for SQ assembly at $\rho=0.3$.} Left: Normalized average expected improvement of GPR predictions. Right: Hypervolume under the Pareto front (black line and inverted triangle) and the fraction of updated potentials on the Pareto front (blue line and triangles). Generations to the left of the dashed vertical line were obtained by global maximization of EI, while generations to the right were obtained by searching for local maxima of EI.}
    \label{fig:sq-pf-convergence}
\end{figure}

\section{Pareto-optimal potentials in design space}
\begin{figure}[H]
    \centering
    \includegraphics[width=\textwidth]{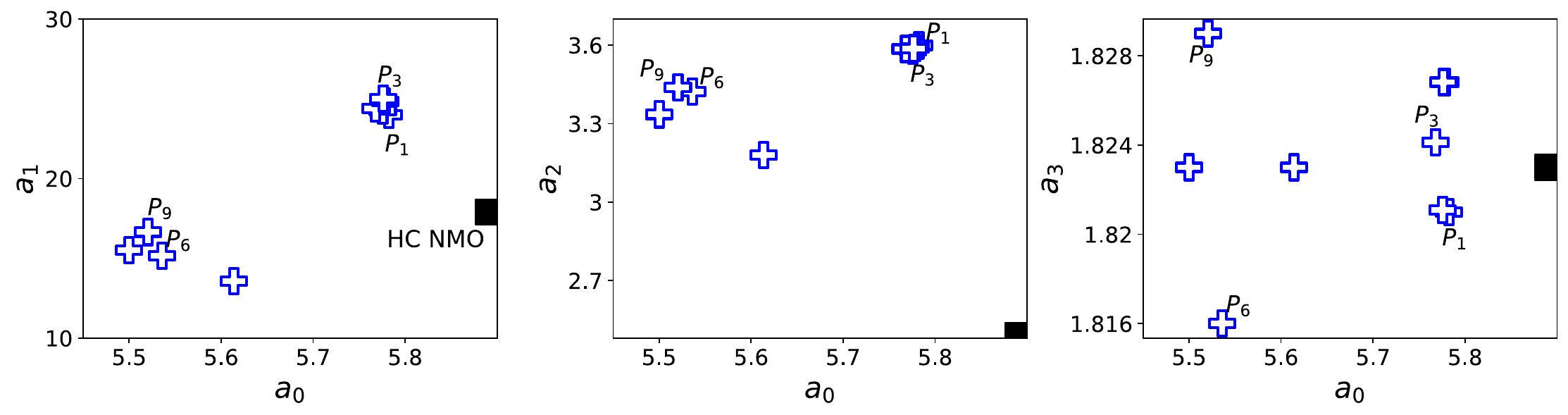}
    \caption{\textbf{Dependence of the Pareto-optimal potentials on the design parameters, $\vec{a}$, for HC self-assembly at $\rho=0.53$ using periodic boundary conditions.}  The Pareto-optimal potentials are not obviously rank-ordered in terms of the design parameters, $\vec{a}$, although they are substantially separated from the HC NMO potential in terms of the parameters $a_0$ and $a_2$.}
    \label{fig:pareto_des_space}
\end{figure}

\figref{fig:pareto_des_space} shows the Pareto-optimal HC potentials at $\rho=0.53$ using periodic boundary conditions, along with the HC NMO potential, in the design space $\vec{a}$.
We find that the Pareto-optimal potentials are not rank-ordered in terms of any individual design parameter or linear combination thereof, likely because the features of the nonlinear pair potential are sensitive to small perturbations in the design parameters.
However, we note that all Pareto-optimal potentials have very different $a_0$ and $a_2$ parameters relative to the HC NMO potential.

\section{Pareto-optimality and potential features for alternative objective definitions and pair potential families}
\begin{figure}[H]
        \centering
        \includegraphics[width=0.7\textwidth]{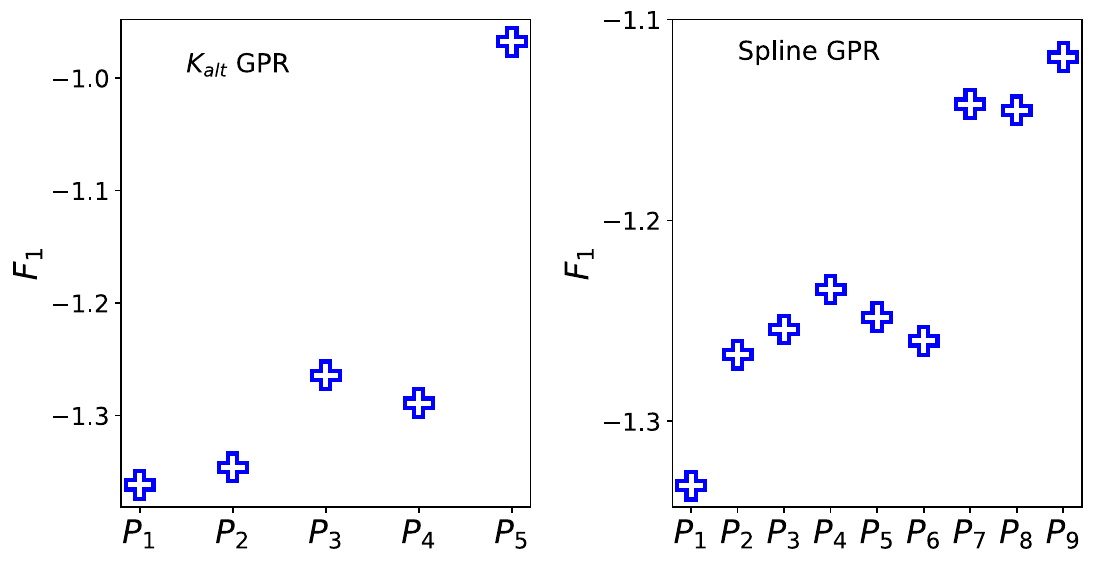}
        \caption{\textbf{Pareto-optimal potentials obtained using alternative objective definitions and pair potential families are also rank-ordered with respect to $F_1$.} Pareto-optimal potentials using (left) $K_{\text{alt}}$ and (right) spline-based potentials for HC assembly at $\rho=0.53$. The right panel shows that the $F_1$ values for the spline-based optimal potentials are slightly lower than those of the original Pareto-optimal potentials (cf.~Figure~6c in the main text), which likely results in the slight under-performance of the spline-based Pareto front relative to the original Pareto front in \figref{fig:spline}.}
        \label{fig:force}
    \end{figure}

\figref{fig:force} shows the rank-ordering of the Pareto-optimal potentials with respect to $F_1$ for the Pareto fronts obtained using the alternate kinetic definition, $K_{\text{alt}}$, and the spline-based potentials for HC assembly at $\rho=0.53$ using periodic boundary conditions.

\section{Learning rate convergence in short trajectory optimization}
\begin{figure}[H]
    \centering
    \includegraphics[width=\textwidth]{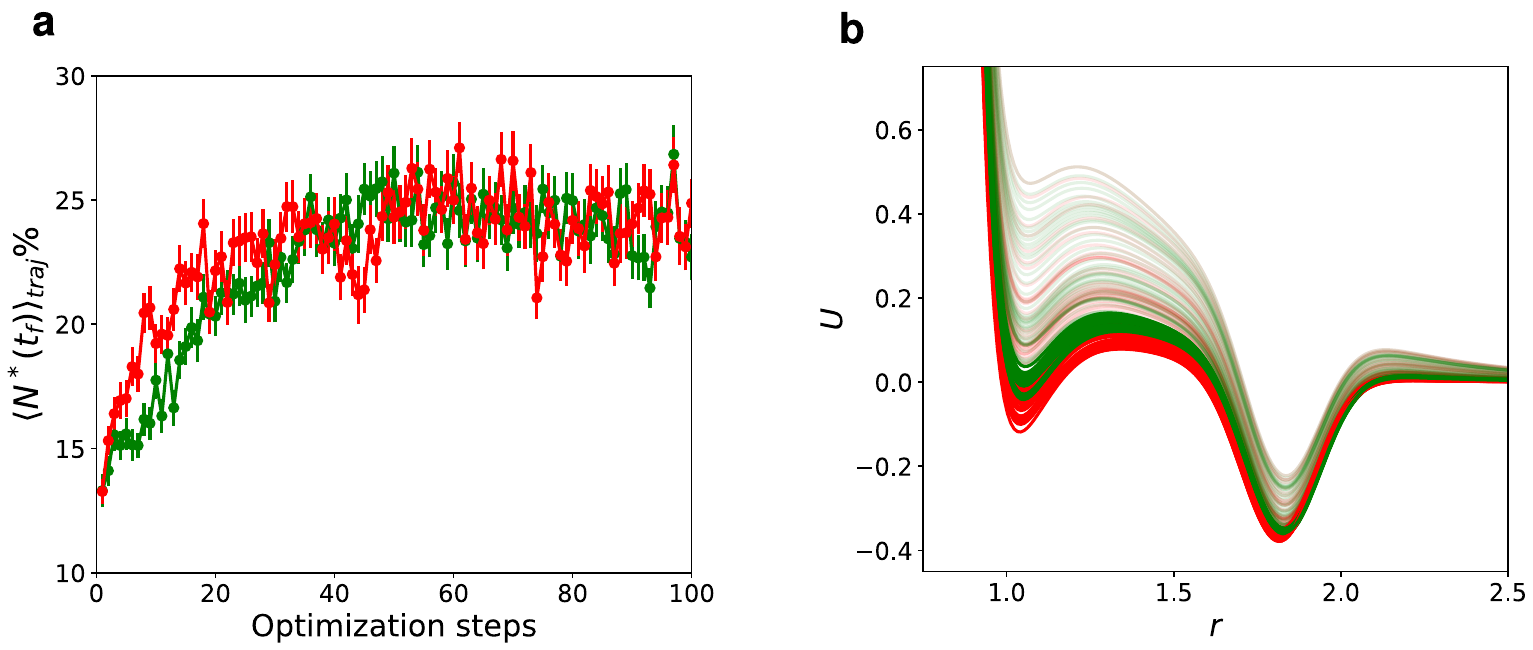}
    \caption{\textbf{Learning rate convergence in trajectory ensemble optimization.} \textbf{a.}~Optimization trajectories showing $\langle N^*(t_\text{f}) \rangle_{\text{traj}}$ for HC self-assembly at $\rho=0.53$ with periodic boundary conditions, using a trajectory duration of $t_\text{f}=400$ and a learning rate of either $l_{\text{rate}}=0.05$ (green) or $l_{\text{rate}}=0.1$ (red). Error bars show the standard error of the mean using $100$ optimization trajectories. \textbf{b.}~Potentials found by the optimization scheme with different learning rates are colored as in \textbf{a}. Intermediate potentials are indicated by transparency, whereas the potentials sampled after optimization step 50 are shown in bold.}
    \label{fig:learning_rate}
\end{figure}

The trajectory ensemble optimization scheme described in Section VI of the main text requires the trajectory duration $t_{\text{f}}$ and learning rate $l_{\text{rate}}$ as hyperparameters.
In \figref{fig:learning_rate}, we show the convergence of $\langle N^*(t_\text{f}) \rangle_{\text{traj}}$ with respect to the learning rate for a fixed $t_\text{f}=400$ for HC self-assembly at $\rho=0.53$ with periodic boundary conditions.
As mentioned in the main text, $N^*$ converges within $\sim 50$ optimization steps, and the trajectories for $l_{\text{rate}}=0.05$ and $l_{\text{rate}}=0.1$ are indistinguishable beyond that point.
At early optimization steps however, the higher learning rate leads to faster convergence, as should be expected.
Also shown are all potentials found in the optimization sequence using the two learning rates.

